\documentclass[a4paper]{jpconf}
\usepackage{graphicx}

\def\H{H\hskip-8.5pt/\hskip2pt}

\def\coeff#1#2{{\textstyle{#1\over #2}}}

\def\VEV#1{\left\langle #1\right\rangle}

\def\lsim{\mathrel{\mathpalette\@versim<}}

\def\gsim{\mathrel{\mathpalette\@versim>}}

\def\Tr{{\rm Tr}\,}

\newcommand{\ba}{\begin{eqnarray}}
\newcommand{\ea}{\end{eqnarray}}

\begin{document}
\title{CPT Violation and Decoherence in Quantum Gravity }

\author{Nick E. Mavromatos}

\address{King's College London, Department of Physics, Strand, London WC2R 2LS, UK}

\ead{nikolaos.mavromatos@kcl.ac.uk}

\begin{abstract}
In this brief review I discuss ways and tests of CPT-Violation in the context of quantum gravity theories with space-time foam vacua, which entail quantum decoherence of matter
propagating in such backgrounds.
I cover a wide variety of sensitive probes, ranging from cosmic neutrinos to meson factories. I pay particular emphasis on associating the latter with specific, probably unique (``smoking-gun''), effects of this type of CPT Violation, related to a modification of Einstein-Podolsky-Rosen (EPR)  correlations in the entangled states of the relevant neutral mesons. I also present some semi-microscopic estimates of these latter effects, in the context of a specific string-inspired model of space-time foam (``D-particle foam'').
\end{abstract}

\section{Introduction: arguments in favour of and against Matter Decoherence in Quantum Gravity}

The theory of Quantum Gravity (QG) is still elusive, despite considerable effort to formulate
mathematically consistent models for the description of quantum fluctuations of space time at
microscopic scales (Planck length scales of order $10^{-35}$ m). An important unknown factor
in such a formulation is the structure of space time at such scales, which could be entirely different from what we are accustomed to from our low-energy experience. In fact, space-time at Planck scales may not even be continuous or commutative.

The traditional approach of considering singular microscopic-scale space-time fluctuations due to the formation and subsequent Hawking evaporation of Planck-size black holes in the vacuum leads to speculations~\cite{hawking} about possible \emph{loss of quantum coherence} (``decoherence'') for matter propagating in such topologically non-trivial space-time vacua. Originally, such properties were thought~\cite{ehns} to be generic features of the space-time foam backgrounds~\cite{wheeler}.

An important issue arising in this context concerns the existence of a well-defined
scattering matrix in the presence of
microscopic black holes (\emph{i.e.} in the regime of strong quantum gravity); the information encoded in matter fields may not be delivered intact to asymptotic observers.
This will lead to an \emph{effective (low-energy) decoherence} of matter fields propagating in foamy backgrounds. Basically, the matter quantum system is an open system, interacting with the ``environment'' of quantum gravitational degrees of freedom, not accessible to low-energy observers performing scattering experiments.

There is a rich phenomenology of such decoherence models, especially in the context of particle physics~\cite{poland}, using sensitive particle probes, such as neutron interferometry, neutral mesons~\cite{ehns,lopez,huet,benatti,fide,msdark} or neutrinos~\cite{lisi,benattineutrino,morgan,icecube,barenboim,bm2,bmsw,jean,rubbia}.
In this context, R.~Wald has proven~\cite{wald} a strong form of violation of CPT symmetry in quantum field theories involving decoherence due to quantum gravity, in the sense that
in such theories the quantum mechanical operator that generates the CPT transformations
is not \emph{well-defined}. In these cases, CPT ``Violation'' refers to a rather \emph{intrinsic} \emph{microscopic time irreversibility} situation (``\emph{intrinsic CPT Violation}''), and it should be disentangled from situations that could characterize some effective field theories of, say, Lorentz violation, \emph{e.g}. of the type encountered in the so-called Standard Model Extension (SME)~\cite{sme}. There, the CPT breaking is associated simply with the non-commutativity of the (well-defined in that case) CPT quantum-mechanical generator with the effective Hamiltonian of the SME.

Subsequent theoretical developments, in particular in the context of string theory,
pointed out to the model dependence of such decoherening effects of QG.
In particular, if \emph{holography} is a property of a string-theory black hole,
either in the more general sense of \cite{thooft}, or in the sense of some form of Anti-De-Sitter/Conformal-Field-Theory (AdS/CFT) correspondence~\cite{maldacena,barbdiscr}, then information may be restored in such space-time foam backgrounds, leading to decoherence-free matter propagation in such stringy space-times.

In fact, inspired by such conjectures, Hawking recently~\cite{hawking2}
retracted his original suggestion~\cite{hawking} of quantum-gravity-induced decoherence by arguing, following Maldacena's treatment of string theory black holes~\cite{malda2}, that in a \emph{Euclidean} path-integral formalism, when one sums up topologically trivial and non-trivial (black-hole) space-times,
any contribution from the latter to the path integral, which would be responsible for the loss of coherence, decays asymptotically in time, leaving \emph{only} the topologically trivial \emph{unitary} space-time contributions. In this way, Hawking argues~\cite{hawking2} that coherence is never lost in quantum fluctuating black holes,
but is \emph{entangled in a holographic way} with the portion of space-time outside the
horizon.  In simple terms, according
to this reasoning, the information is not lost but may be so mangled that
it cannot be easily extracted by an asymptotic observer. Hawking drew the analogy
to information encrypted in ``a burnt out encyclopedia'', where the
information is radiated away in the environment, but there is no paradox,
despite the fact that it is impossibly difficult to recover.
In such a holographic Universe one may therefore expect quantum coherence and hence CPT to be exact properties of the quantum gravity interactions.

However, there are fundamental issues we consider as unanswered by the above
interesting arguments.  On
the technical side, one issue that causes concern is the Euclidean
formulation of QG. According to Hawking this is the only sensible way to
perform the path integral over geometries. However, given the uncertainties
in analytic continuation, it may be problematic. Additionally, it has been
argued ~\cite{hawking2} that the dynamics of formation and evaporation
of (microscopic) black holes is unitary using Maldacena's holographic
conjecture of AdS/CFT correspondence~\cite{maldacena} for the case of
Anti-de-Sitter (supersymmetric) space-times. This framework describes the
process in a very specific category of foam, and may not be valid generally
for theories of QG. Even in this context, though, the r\^{o}le of the
different topological configurations is actually important, a point recently
emphasised by Einhorn~\cite{einhorn}. In Maldacena's treatment of
black holes~\cite{malda2}, the non-vanishing of the contributions to the
correlation functions due to the topologically non-trivial configurations is
required by unitarity. Although such contributions vanish in semiclassical
approximations, the situation may be different in the full quantum theory,
where the r\^{o}le of stretched and fuzzy (fluctuating) horizons may be
important, as pointed out by Barbon and Rabinovici~\cite{barbon}.

The information paradox is acutest~\cite{einhorn} in the case of
gravitational collapse to a black hole from a pure quantum mechanical state,
without a horizon; the subsequent evaporation due to the celebrated
Hawking-radiation process, leaves an apparently ``thermal'' state. It is in
this sense that the analogy~\cite{hawking2} is made with the encoding
of information in the radiation of a burning encyclopedia. However the
mangled form of information in the burnt out encyclopedia, is precisely the
result of an interaction of the encyclopedia with a heat bath that burned
its pages, thereby leading to an \textit{irreversible} process. The
information cannot be retrieved due to entropy production in the process.

In our view, if microscopic black holes, or other defects forming space-time
foam, exist in the vacuum state of quantum gravity (QG), this state will
constitute an ``environment'' which will be characterized by some \textit{%
entanglement entropy}, due to its interaction with low-energy
matter.  Within
critical string theory, arguments have been given that entanglement
entropy can characterise the number of microstates of Anti-de-Sitter
black holes~\cite{entanglement}, but we do not find these to give a complete answer to the issue.
Indeed, a space-time foam situation in a theory of quantum gravity, in particular
 the formation and subsequent evaporation of a (microscopic) black hole,
may correspond to a non-equilibrium process in string theory, being described by non-critical
string theory (i.e. string models, which in their perturbative $\sigma$-model
treatment correspond to models involving non-conformal world-sheet deformations, dressed by the so-called Liouville mode~\cite{ddk}). Such non-critical strings, with the Liouville mode identified (dynamically)
with an \emph{irreversible} target time variable, constitute sorts of \emph{non-equilibrium} string theory~\cite{emn}.  In this latter sense, the whole issue of holography or quantum coherence of matter may be at stake.

In this approach, there are concrete stringy models of space-time foam~\cite{emw},
inspired by the modern approach to string theory, involving space-time
solitonic structures, the so-called Dirichlet-(D-)branes~\cite{polchinski}.
According to this approach, our Universe might be such a three-brane, propagating in a higher dimensional bulk space. In the space-time foam model of \cite{emw}, for instance, the bulk space is punctured by massive (with masses or order of the string scale $M_s$) point-like D-branes (termed D-particles), whose topologically non-trivial interactions with strings attached on the three -brane Universe, representing ordinary matter excitations, lead to
decoherening, time-irreversible  effects. Such effects involve the splitting of the matter strings by the D-particles, and the recoil of the latter, as a result of momentum conservation. The recoil of the massive defect is a process that can be described in the context of non-critical string theory~\cite{emn,emw}, as a result of back reaction effects of these massive objects on the space time background. Such processes lead to local distortions of the space time, which upon quantization
lead to stochastic space-time fluctuations (``D-particle foam'').

As a result of such stochastic processes, the space-time fluctuations, due to the recoil of the D-particle defect, carry the relevant information, and in this way the full string theory is unitary. Nevertheless,
as argued in \cite{emw,msdark} this information cannot be recovered by local scattering experiments,
and hence the effective low energy theory ``appears'' non unitary. This leads to an effective decoherence, with observable (in principle) effects, for instance damping signatures in several particle processes, such as particle flavour oscillations (neutrinos, neutral mesons \emph{etc}.) This approach has been followed by the
author and collaborators~\cite{poland,msdark} in many
recent phenomenological tests of such microscopic stringy models of space-time foam.

Moreover, the low-energy CPT operator may appear in these models as ill-defined, following the arguments of \cite{wald}, leading to some specific signatures in entangled states of neutral mesons, that may be unique for this type of quantum gravity induced decoherence. Such signatures involve modifications of the Einstein-Podolsky-Rosen (EPR) particle correlators in entangled states of neutral mesons, an effect
proposed in \cite{bmp} ($\omega$-effect) as a probably unique signature of the ill-defined nature of CPT operator in decoherent effective theories of quantum gravity. In fact, explicit estimates of this effects in the context of the D-particle space-time foam model~\cite{emw} have been given in \cite{bms},
pointing to the possibility that such effects may be falsifiable already at the next-generation meson factories, such as an upgrade in the DA$\Phi$NE detector~\cite{adidomenico}.

The above arguments suggest that the situation associated with the issue of
unitarity of effective matter theories in foamy space-times remains unresolved, and in this sense
experimental searches for (and phenomenological models of) quantum decoherence and induced CPT symmetry violation in QG, using sensitive probes, should be performed. After all, it is the Experiment that points out to the right track for our understanding of a physical theory, and the still elusive theory of quantum gravity may not be an exception to that rule.

It is the point of this article to review briefly the situation with CPT Violation in some models of quantum gravity entailing decoherence, and discuss the pertinent effects in a variety of particle physics probes, ranging from cosmic neutrinos to the above-mentioned neutral meson facilities. As part of the discussion, I will also review the order of magnitude estimates for such effects in various models, with the purpose of emphasizing the delicate model dependence of such estimates. This is an important aspect to be taken into account in searches for such effects
in various facilities.

The structure of the talk is as follows: in the next section \ref{sec:2}, I discuss briefly the relevant formalism to be used when discussing QG-induced decoherence of matter in various theoretical models, and present some order of magnitude estimates of the relevant effects. As we shall see, some of these models are very far from being falsifiable in the foreseeable future, while others can be put to experimental test in upcoming facilities. In section \ref{sec:3}, I discuss the phenomenology of QG-decoherence in neutrino oscillations and give the current bounds for the relevant parameters, while in section \ref{sec:4}, I present the relevant phenomenology in the case of neutral mesons. I pay particular attention to tests performed in meson factories, involving entangled states, where modifications of the Einstein-Podolsky-Rosen (EPR) correlations ($\omega$-effect)
may appear as a result of the ill-defined nature of the CPT operator in QG decoherent models. I discuss the $\omega$-effect in both kaon and B-meson factories~\cite{bomega}, with the purpose of describing probably unique (``smoking-gun'') effects of QG decoherence  as far as this type of CPT Violation is concerned. I also give some estimates of the
effect within the context of the afore-mentioned model of D-particle foam.
Conclusions and outlook are presented in section \ref{sec:5}.

\section{Quantum Gravity Decoherence and intrinsic CPT Violation: concepts, properties and formalism \label{sec:2}}

\subsection{Ill-defined CPT and Decoherence?}

I commence the discussion with an important issue concerning the possible fate of C(harge)P(arity)T(time-reversal) Symmetry in theories of Quantum Gravity, entailing decoherence of
matter.
In the presence of Quantum Gravity (QG) fluctuations the preservation of CPT symmetry is by no means guaranteed. Indeed,  CPT invariance is ensured in \emph{flat space-times} by a theorem
applicable to any local relativistic quantum field theory of the type used
to describe currently the standard phenomenology of particle physics.
More precisely the \emph{ CPT theorem} states~\cite{cpt}:
\emph{Any quantum theory formulated on flat space-times is symmetric
under the combined action of CPT transformations, provided the
theory respects (i) Locality, (ii) Unitarity (i.e. conservation of
probability) and (iii) Lorentz invariance.}

The validity of any such theorem in the QG regime is an open and challenging issue since it is linked with our understanding of the nature of space-time at
(microscopic) Planckian distances $10^{-35}$~m. However there are reasons to believe that the CPT theorem \emph{may not} be valid (at least
in its strong form) in highly curved space-times with event horizons, such as those in the vicinity of
black holes, or more generally in some QG models involving
{\it quantum space-time foam} backgrounds~\cite{wheeler}. The latter
are characterized by singular quantum fluctuations of space-time
geometry, such as microscopic black holes, {\it etc.}, with event horizons of
microscopic Planckian size. Such backgrounds result in {\it
apparent} violations of {\it unitarity} in the following sense:
there is some part of the initial information (quantum numbers of
incoming matter) which ``disappears'' inside the microscopic event
horizons, so that an observer at asymptotic infinity will have to
trace over such ``trapped'' degrees of freedom. One faces therefore
a situation where an initially pure state evolves in time and
becomes mixed. The asymptotic states are described by density
matrices, defined as
\begin{equation}
\rho _{\rm out} = {\rm Tr}_{M} |\psi ><\psi|~, \end{equation}
where
the trace is over trapped (unobserved) quantum states that
disappeared inside the microscopic event horizons in the foam. Such
a non-unitary evolution makes it impossible to define a
standard quantum-mechanical scattering matrix. In ordinary local
quantum field theory, the latter connects asymptotic state vectors
in a scattering process
\begin{equation}
|{\rm out}> = S~|{\rm in}>,~S=e^{iH(t_f - t_i)}~,
\end{equation}
where $t_f - t_i$ is the duration of the scattering (assumed to be
much longer than other time scales in the problem, i.e.
${\rm lim}~t_i \to -\infty$, $t_f \to +\infty$). Instead, in
foamy situations, one can only define an operator that connects
asymptotic density matrices~\cite{hawking}:
\begin{equation}\label{dollar}
\rho_{\rm out} \equiv {\rm Tr}_{M}| {\rm out} ><{\rm out} | = \$
~\rho_{\rm in},~ \quad \$ \ne S~S^\dagger.
\end{equation}
The lack of factorization is attributed to the \emph{apparent} loss of
unitarity of the effective low-energy theory, defined as the part of
the theory accessible to low-energy observers performing scattering
experiments. In such  situations particle phenomenology has to be
reformulated~\cite{ehns,poland} by viewing our low-energy world as
an open quantum system and using (\ref{dollar}). Correspondingly,
the usual Hamiltonian evolution of the wave function is replaced
by the Liouville equation for the density matrix~\cite{ehns}
\begin{equation}\label{evoleq}
\partial_t \rho = i[\rho, H] + \delta\H \rho~,
\end{equation}
where $\delta\H \rho $ is a correction of the form normally found
in open quantum-mechanical systems~\cite{lindblad}, although more general
forms are to be expected in QG~\cite{msdark}.
This is what we denote by QG-induced {\it decoherence}, since the interaction with the quantum-gravitational environment  results in
quantum decoherence of the matter system, as is the case of open quantum mechanical systems in general~\cite{zurek,kiefer}.

The \$ matrix is {\it not invertible}, and this reflects the
effective unitarity loss. Since one of the requirements of CPT
theorem ( viz. unitarity) is violated there is no
CPT invariance (in theories which have CPT invariance in the absence of trapped states). This semi-classical analysis  leads to
more than a mere violation of the symmetry. The CPT
operator itself is \emph{not well-defined}, at least from an
effective field theory point of view. This is a strong form of CPT
violation (CPTV)~\cite{wald} and can be summarised by: \emph{In an open (effective) quantum
theory, interacting with an environment, e.g., quantum
gravitational, where} $ \$ \ne SS^\dagger $, \emph{CPT invariance is
violated, at least in its strong form}.
This form of violation introduces a fundamental arrow of
time/microscopic time irreversibility, unrelated in principle to CP
properties. Such
decoherence-induced CPT Violation (CPTV) should occur in effective field
theories, since the low-energy experimenters do not have access to
all the degrees of freedom of QG (e.g., back-reaction
effects, \emph{etc.}). Some have conjectured that full CPT invariance
could be restored in the (still elusive) complete theory of QG.
In such a case, however, there may be~\cite{wald}
a \emph{weak form of CPT invariance}, in the sense of the possible existence
of \emph{decoherence-free subspaces} in the space of states of a
matter system. If this situation is realized, then the strong form
of CPTV will not show up in any measurable quantity (that
is, scattering amplitudes, probabilities \emph{etc.}).

The weak form of CPT invariance may be stated as follows: \emph{Let}
$\psi \in {\cal H}_{\it in}$, $\phi \in {\cal H}_{\it out}$
\emph{denote pure states in the respective Hilbert spaces ${\cal H}$
of in and out states, assumed accessible to experiment. If
$\theta $ denotes the (anti-unitary) CPT operator acting on pure
state vectors, then weak CPT invariance implies the following
equality between transition probabilities}
\begin{equation}
{\cal P}(\psi \to \phi) = {\cal P}(\theta^{-1}\phi \to \theta
\psi)~. \label{weakcpt}
\end{equation}
Experimentally, at least in principle,  it is possible to test
equations such as (\ref{weakcpt}), in the sense that, if decoherence
occurs, it induces (among other modifications) damping factors
in the time profiles of the corresponding transition probabilities.
The diverse experimental techniques for testing decoherence
range from
terrestrial laboratory experiments (in high-energy, atomic and
nuclear physics) to astrophysical observations of light from distant
extragalactic sources and high-energy cosmic neutrinos~\cite{poland}.

\subsection{Models of Quantum-Gravity Decoherence}

The models of quantum gravity inducing decoherence can be divided in roughly two
categories, as far as the effects in the propagation of matter are concerned:
\begin{itemize}
\item{(i)}  models in which the environmental decoherence effects can be cast in a Hamiltonian form, affecting only the form of the Hamiltonian of the matter system during its propagation in the space-time background. In such models, back-reaction effects onto the space time, for which the concept of the
local effective lagrangian may break down, are ignored. In such a category lie the models of stochastically fluctuating geometries, which we discuss in the next subsection.

\item{(ii)} Models in which the propagation of matter entail a \emph{breakdown of the local effective lagrangian formalism}, and with it the concept of the scattering matrix, as we discussed briefly above. In such cases, back reaction effects onto the structure of space time, as a result of topologically non trivial interactions of matter with singular space-time fluctuations or defects, result in non-Hamiltonian terms appearing in the evolution of the reduced density matrix of matter. The resulting formalism is that of open quantum mechanical systems with an environment.
    The most commonly used phenomenological approach is that due to Lindblad~\cite{lindblad}, which we discuss in subsequent subsections of this article.

\end{itemize}

The above division is of course somewhat artificial, and in fact in complete models of quantum gravity one expects both effects to be simultaneously present. However, for instructive purposes we discuss the relevant phenomenology separately.

\subsubsection{Stochastically Fluctuating Geometries, Light Cone
Fluctuations and Decoherence}

\begin{figure}[htb]
\begin{center}
  \includegraphics[width=0.4\textwidth]{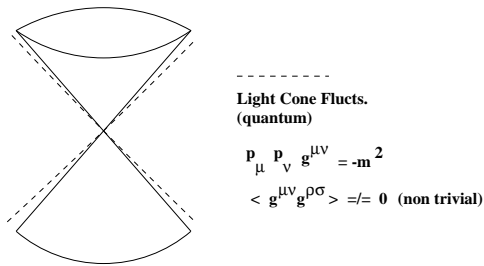}
\end{center}
\caption{In stochastic space-time models of QG the
light cone may fluctuate, leading to decoherence and quantum
fluctuations of the speed of light in ``vacuo''.} \label{lcf}
\end{figure}

If the ground state of QG consists of ``fuzzy'' space-time, i.e.,
stochastically-fluctuating metrics, then a plethora of interesting
phenomena may occur, including light-cone
fluctuations~\cite{ford,emn} (c.f. Fig.~\ref{lcf}).
Such effects will lead to stochastic fluctuations in, say, arrival
times of photons with common energy, which can be detected with
high precision in astrophysical experiments~\cite{efmmn,ford}.
In addition, they may give rise to
decoherence of matter, in the sense of induced time-dependent
damping factors in the evolution equations of the (reduced) density
matrix of matter fields~\cite{emn,msdark}.

Such ``fuzzy'' space-times are formally represented by metric
deviations which are fluctuating randomly about, say, flat Minkowski
space-time: $g_{\mu\nu} = \eta_{\mu\nu} + h_{\mu\nu}$, with $\langle
\cdots \rangle$ denoting statistical quantum averaging, and $\langle
g_{\mu\nu} \rangle = \eta_{\mu\nu} $  but $\langle h_{\mu\nu}(x)
h_{\lambda \sigma}(x') \rangle \ne 0 $,  i.e., one has only quantum
(light cone) fluctuations but not mean-field effects on dispersion
relations of matter probes. In such a situation Lorentz symmetry is
respected on the average, \emph{but} not in individual measurements.

The path of light follows null geodesics $ 0 = ds^2 =
g_{\mu\nu}dx^\mu dx^\nu $, with non-trivial fluctuations in geodesic
deviations, ${D^2n^{\mu}\over D\tau^2} =
-R^{\mu}_{\alpha\nu\beta}u^{\alpha}n^{\nu}u^{\beta}\,;$ in a
standard general-relativistic notation, $D/D\tau$ denotes the
appropriate covariant derivative operation,
$R^{\mu}_{\alpha\nu\beta}$ the (fluctuating) Riemann curvature
tensor, and $u^\mu$ ($n^\mu$) the tangential (normal) vector along
the geodesic.

Such an effect causes primarily fluctuations in the arrival time of
photons at the detector ($|\phi \rangle$=state of gravitons, $|0
\rangle$= vacuum state) {\small $$
 \Delta t_{obs}^2=|\Delta t_{\phi}^2-\Delta t_0^2 |=
{|\langle \phi| \sigma_1^2 |\phi\rangle-\langle 0| \sigma_1^2
|0\rangle|\over r^2}\equiv {|\langle \sigma_1^2 \rangle_R|\over r}\,
,
$$}
where {\small \begin{eqnarray}  && \langle \sigma_1^2 \rangle_R
=\frac{1}{8}(\Delta r)^2 \int_{r_0}^{r_1} dr \int_{r_0}^{r_1} dr'
\:\,
n^{\mu} n^{\nu} n^{\rho} n^{\sigma} \:\, \nonumber \\
&& \langle \phi| h_{\mu\nu}(x) h_{\rho\sigma}(x')+
 h_{\mu\nu}(x') h_{\rho\sigma}(x)|\phi \rangle \label{variance} \end{eqnarray}}and
the two-point function of graviton fluctuations can be evaluated
using standard field theory techniques~\cite{ford}.

Apart from the stochastic metric fluctuations, however, the aforementioned
effects could also
induce decoherence of matter propagating in these types of
backgrounds~\cite{msdark,jean}, a possibility of particular interest for the
purposes of the present article.
Through the theorem of Wald~\cite{wald}, this implies that
the CPT operator may not be well-defined, and hence one also has a
breaking of CPT symmetry in such cases.

The most important effect of the associated decoherence will be damping
in, say, oscillation probabilities during flavoured particle propagation in such
fuzzy space times.
This problem has been analysed in some detail in \cite{msdark,jean} and we only state here the
final formula for fermion oscillation. We consider for completeness Robertson-Walker space times,
with scale factors $a(z)$, with $z$ the red-shfit,
but ignore for our purposes particle production, which is a case relevant for late-times in the Universe evolution, as appropriate for cosmic particle phenomenology. The stochastically fluctuating (inverse) metric is taken to be~\cite{jean}:
\ba\label{fluctuatingmetric}
g^{\mu\nu} = a^{-2}(z)\left(\eta^{\mu\nu}+h^{\mu\nu}\right),
\ea
with $\eta^{\mu\nu}$ a Minkowski metric
and the stochastic fluctuations components $h^{\mu\nu}$ are assumed (for brevity) coordinate-independent and much small than one in absolute value.

The relevant
stochastic space-time oscillation probability for oscillations between the flavours $\alpha , \beta$, averaged over Gaussian (for concreteness) space-time fluctuations, with a (renormalised) variance $\sigma^2 \equiv \langle \sigma_1^2 \rangle_R$ (c.f. (\ref{variance})),
reads~\cite{jean}:
\begin{eqnarray}
\label{Pabaverage}
<P_{\alpha\to\beta}>&=&\left(\frac{1}{1+z}\right)^2\exp\left\{i\left(\frac{m_i^2-m_j^2}{2k^2}
-\frac{\sigma^2}{16}\frac{m_j^4-m_i^4}{k^4}
\frac{z(z +2)}{(1+z)^2}\right)I(z)\right\} \times  \nonumber \\
&&\times\exp\left\{-\frac{\sigma^2}{32}\left(\frac{m_i^2-m_j^2}{k^2}\right)^2I^2(z)\right\}\nonumber,
\end{eqnarray}
where $k \equiv |\vec k|$ is the magnitude of the spatial momentum of the probe,
$m_i$ denote masses of the mass eigenstates, $z$ is the cosmic redshift, and the factor
\begin{equation}
I(z)= k\int_{0}^z \frac{dz'}{H(z')}\frac{1}{(1+z')^2}.
\end{equation}
where $H(z)$ is the Hubble parameter, which in a standard $\Lambda$CDM spatially-flat Cosmology, assumes the form: $H(z)  = H_0 \sqrt{\Omega_\Lambda + \Omega_{{\rm M}} (1 + z)^3}$, with $H_0$
the present-epoch value and $\Omega_\Lambda$ $(\Omega_M)$ the present-day cosmological constant (matter, including dark matter) contribution to the vacuum energy density of the Universe, in units
of the critical density. In Expression (\ref{Pabaverage}), the first line corresponds to the oscillation term in a smooth expanding Universe, the second line leads to a correction to the phase of these oscillations, arising from the stochastic metric fluctuations,
and the third line represents a \emph{damping}, also generated by the random fluctuations of the metric, which, as mentioned previously, is a characteristic effect of decoherent theories.
Notice for later use that the damping in the Gaussian model is proportional to the exponential of the square of the time, ${\rm exp}[-\left(\dots \right)t^2]$.

In fact, it is worthy of mentioning that if one ignores temporal derivatives of the scale factor,
and hence particle production in the expanding Universe, then the expressions for the oscillation probability between flavoured fermions and bosons are similar.

\subsubsection{Lindblad  Decoherence - Environmental non-Hamiltonian Evolution}

If the effects of the environment are such
that the modified evolution equation of the (reduced) density matrix
of matter $\rho$~\cite{kiefer} is linear, one can write
down a Lindblad evolution equation~\cite{lindblad}, provided that
(i) there is (complete) positivity of $\rho$, so that negative
probabilities do not arise at any stage of the evolution, (ii) the
energy of the matter system is conserved on the average, and (iii)
the entropy is increasing monotonically.

For $N$-level systems, the generic decohering Lindblad evolution for $\rho$ reads
\begin{eqnarray}\label{lindevol}
\frac{\partial \rho_\mu}{\partial t} = \sum_{ij} h_i\rho_j
{f}_{ij\mu} + \sum_{\nu} {L}_{\mu\nu}\rho_\nu~,\quad~\mu, \nu = 0, \dots N^2 -1, \quad i,j = 1, \dots N^2 -1~,
\end{eqnarray}
where the $h_i$ are Hamiltonian terms, expanded in an appropriate
basis, and the decoherence matrix $L$ has the form:
\begin{equation}
{L}_{0\mu}={L}_{\mu 0} =0~, \quad {L}_{ij} = \frac{1}{4}\sum_{k,\ell
,m} c_{l\ell}\left(-f_{i\ell m}f_{kmj} + f_{k i m}f_{\ell m
j}\right)~,
\label{cmatrix}
\end{equation}
 with $c_{ij}$ a {\it positive-definite matrix} and $f_{ijk}$
the structure constants of the appropriate $SU(N)$ group.
In this generic phenomenological description of decoherence, the
elements ${L}_{\mu\nu}$ are free parameters,
to be determined by experiment. We shall come back to this point in
the next subsection, where we discuss neutral kaon decays.

A rather characteristic feature of this equation is the
appearance of exponential damping, $e^{-(...)t}$, in interference terms
of the pertinent quantities (for instance, matrix elements $\rho$,
or asymmetries in the case of the kaon system, see below).
The exponents are proportional to (linear
combinations) of the elements of the decoherence
matrix~\cite{lindblad,ehns,kiefer}.

A specific example of Lindblad evolution is the propagation
of a probe in a medium with a stochastically fluctuating density~\cite{loreti}.
The formalism can be adapted to the case of stochastic space-time foam~\cite{msdark,bmsw}.

The stochasticity of the space-time foam medium
is best described~\cite{bmsw} by
including in the time evolution of the neutrino
density matrix a a time-reversal (CPT) breaking
decoherence matrix of a double commutator form~\cite{loreti,bmsw},
\begin{eqnarray}
\partial_t \langle \rho\rangle =  L[\rho]~, \quad
L[\rho]=
-i[H + H'_{I},\langle \rho\rangle]-\Omega^2[H'_I,[H'_I,\langle \rho\rangle]]
\label{double}
\end{eqnarray}
where $\langle n(r) n(r') \rangle = \Omega^2n_0^2
\delta (r - r') $ denote the stochastic (Gaussian) fluctuations of
the density of the medium,
and
 \ba
H'_I=\left(%
\begin{array}{cc}
  (a_{\nu_{e}}-a_{\nu_{\mu}})\cos^2(\theta)  & (a_{\nu_{e}}-a_{\nu_{\mu}})\frac{\sin2\theta}{2}   \\
  (a_{\nu_{e}}-a_{\nu_{\mu}})\frac{\sin2\theta}{2}  & (a_{\nu_{e}}-a_{\nu_{\mu}})\sin^2(\theta)  \\
\end{array}%
\right) \ea
is the MSW-like interaction~\cite{msw}
in the mass eigenstate basis, where
$\theta$ is the mixing angle.
This double-commutator decoherence is a
specific case of Lindblad evolution \ref{lindevol}
which guarantees complete positivity of the time evolved
density matrix. For gravitationally-induced MSW effects (due to, say, black-hole foam
models as in \cite{bm2,msdark}), one may denote the difference, between neutrino flavours,
of the effective interaction strengths, $a_i$,
 with the environment by:
\ba
\Delta a_{e\mu} \equiv a_{\nu_e}-a_{\nu_\mu} \propto G_N n_0
\ea
with $G_N=1/M_P^2$, $M_P \sim 10^{19}~{\rm GeV}$, the four-dimensional
Planck scale, and
in the case of
the gravitational MSW-like effect~\cite{bm2} $n_0$
represents the
density  of charge black hole/anti-black hole pairs.
This gravitational coupling replaces the weak interaction
Fermi coupling constant $G_F$ in the conventional MSW effect~\cite{msw}.

For two generation neutrino models, the corresponding oscillation probability $\nu_e \leftrightarrow
\nu_\mu$ obtained from (\ref{double}), in the small
parameter $\Omega^2 \ll 1$, which we assume here, as appropriate for the weakness of gravity fluctuations, reads to leading order:
{\small
\begin{eqnarray}
&&    P_{\nu_e\to \nu_{\mu}}=   \nonumber \\
    && \frac{1}{2} + e^{-\Delta a_{e\mu}^2\Omega^2t(1+\frac{\Delta_{12}^2}{4\Gamma}
(\cos(4\theta)-1))}
    \sin(t\sqrt{\Gamma})\sin^2(2\theta)\Delta
a_{e\mu}^2\Omega^2\Delta_{12}^2
    \left(\frac{3\sin^2(2\theta)\Delta_{12}^2}{4\Gamma^{5/2}}
-\frac{1}{\Gamma^{3/2}}\right) \nonumber \\
    && -e^{-\Delta
    a_{e\mu}^2\Omega^2t(1+\frac{\Delta_{12}^2}{4\Gamma}(\cos(4\theta)-1))}
    \cos(t\sqrt{\Gamma})
\sin^2(2\theta)\frac{\Delta_{12}^2}{2\Gamma}  \nonumber \\
    &&-e^{-\frac{\Delta a_{e\mu}^2\Omega^2 t \Delta_{12}^2\sin^2(2\theta)}{\Gamma}}
    \frac{(\Delta a_{e\mu}+\cos(2\theta)\Delta_{12})^2}{2\Gamma}
\label{2genprob}
\end{eqnarray}}
where $\Gamma= (\Delta a_{e\mu}\cos(2\theta)+\Delta_{12})^2+\Delta
a_{e\mu}^2\sin^2(2\theta)~,$ $\Delta_{12}=\frac{\Delta m_{12}^2}{2p}~.$

{}From (\ref{2genprob}) we easily conclude
that the exponents of the
damping factors due to the stochastic-medium-induced decoherence,
are of the generic form, for $t = L$, with $L$ the oscillation
length (in units of $c=1$):
\ba
{\rm exponent} \sim
-\Delta a_{e\mu}^2\Omega^2 t f(\theta)~;~
f(\theta) =
1+\frac{\Delta_{12}^2}{4\Gamma}(\cos(4\theta)-1)~, ~{\rm or} ~
\frac{\Delta_{12}^2\sin^2(2\theta)}{\Gamma}
\label{gammadelta}
\ea
that is proportional to the stochastic fluctuations of the
density of the medium.
The reader should note at this stage that, in
the limit $\Delta_{12}\to 0$, which could characterise the situation
in \cite{bm2}, where the space-time foam effects on the
induced neutrino mass difference are the dominant ones, the damping
factor is of the form $ {\rm exponent}_{{\rm gravitational~MSW}}
\sim -\Omega^2 (\Delta a_{e\mu})^2 L~,$ with the precise value of the
mixing angle $\theta$ not affecting the leading order of the various
exponents. However, in that case, as follows from (\ref{2genprob}),
the overall oscillation probability is suppressed by factors
proportional to $\Delta_{12}^2 $, and, hence, the stochastic
gravitational MSW effect~\cite{bm2}, although in principle
capable of inducing mass differences for neutrinos, however does not
suffice to produce the bulk of the oscillation probability, which is
thus attributed to conventional flavour physics.

\subsubsection{Non-critical string-framework for decoherence}

A different type of decoherence in quantum gravity, which goes beyond Lindblad, is provided by the so-called non-critical
string approach to quantum  space-time foam, a sort of non-equilibrium approach advocated in \cite{emn}.
In this subsection we review briefly the relevant formalism, as it provides~\cite{bmp} an explicit example for the kind of CPT ill-defined nature advocated in \cite{wald}.

String theory is to date the most consistent theory of quantum gravity.
In first-quantized string theory, de Sitter or other space-time backgrounds with space-time horizons, such as microscopic black holes of the type encountered in a space-time foam background, are {\it not conformal} on the world-sheet. The corresponding $\beta$-functions are non vanishing. Such non conformal backgrounds can be rendered conformal upon dressing the theory with the so-called Liouville mode~\cite{ddk}. The resulting $\sigma$-model world-sheet theory, then,
contains an extra target-space coordinate. In cases where the deviation from
conformality is supercritical, the Liouville mode has a time-like signature, and can be identified with the target time~\cite{emn}. This formalism
provides a mathematically consistent way of incorporating de Sitter and
other backgrounds with cosmological horizons in string theory.

It can be shown, that the propagation of string matter in such non-conformal backgrounds is decoherent, the decoherence term being proportional to the world-sheet $\beta$-function. The following master equation for the evolution of stringy low-energy matter
in a non-conformal $\sigma $-model~can be derived\cite{emn}
\begin{equation}
\partial _{t}\rho =i\left[ \rho ,H\right] +:\beta ^{i}{\cal G}_{ij}\left[
g^{j},\rho \right] :  \label{master}
\end{equation}%
where $t$ denotes time (Liouville zero mode), the $H$ is the effective
low-energy matter Hamiltonian, $g^{i}$ are the quantum background target
space fields, $\beta ^{i}$ are the corresponding renormalization group $%
\beta $ functions for scaling under Liouville dressings and ${\cal G}_{ij}$
is the Zamolodchikov metric \cite{zam,kutasov} in the moduli space of the
string. To lowest order in the background field expansion the double colon symbol in (\ref{master}) represents the operator
ordering $:AB:=\left[ A,B\right] $ . The index $i$ labels the different
background fields as well as space-time. Hence the summation over $i,j$ in (%
\ref{master}) corresponds to a discrete summation as well as a covariant
integration $\int d^{D+1}y\,\sqrt{-g}$\bigskip\ where $y$ denotes a set of $%
\left( D+1\right) $-dimensional target space-time co-ordinates and $D$ is
the space-time dimensionality of the original non-critical string.

\begin{figure}[htb]
\begin{center}
\includegraphics[width=0.5\textwidth]{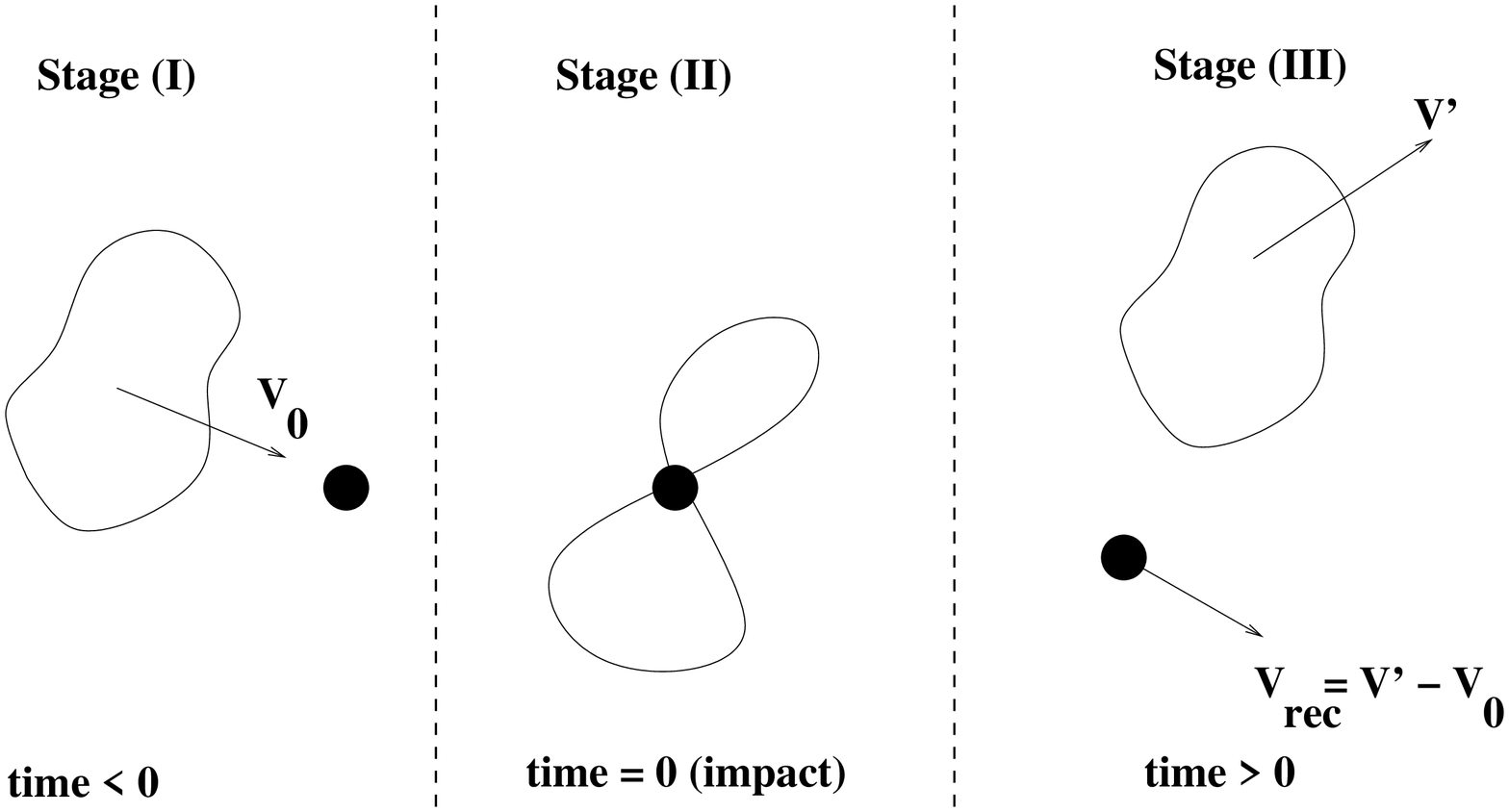}
\hfill \includegraphics[width=0.4\textwidth]{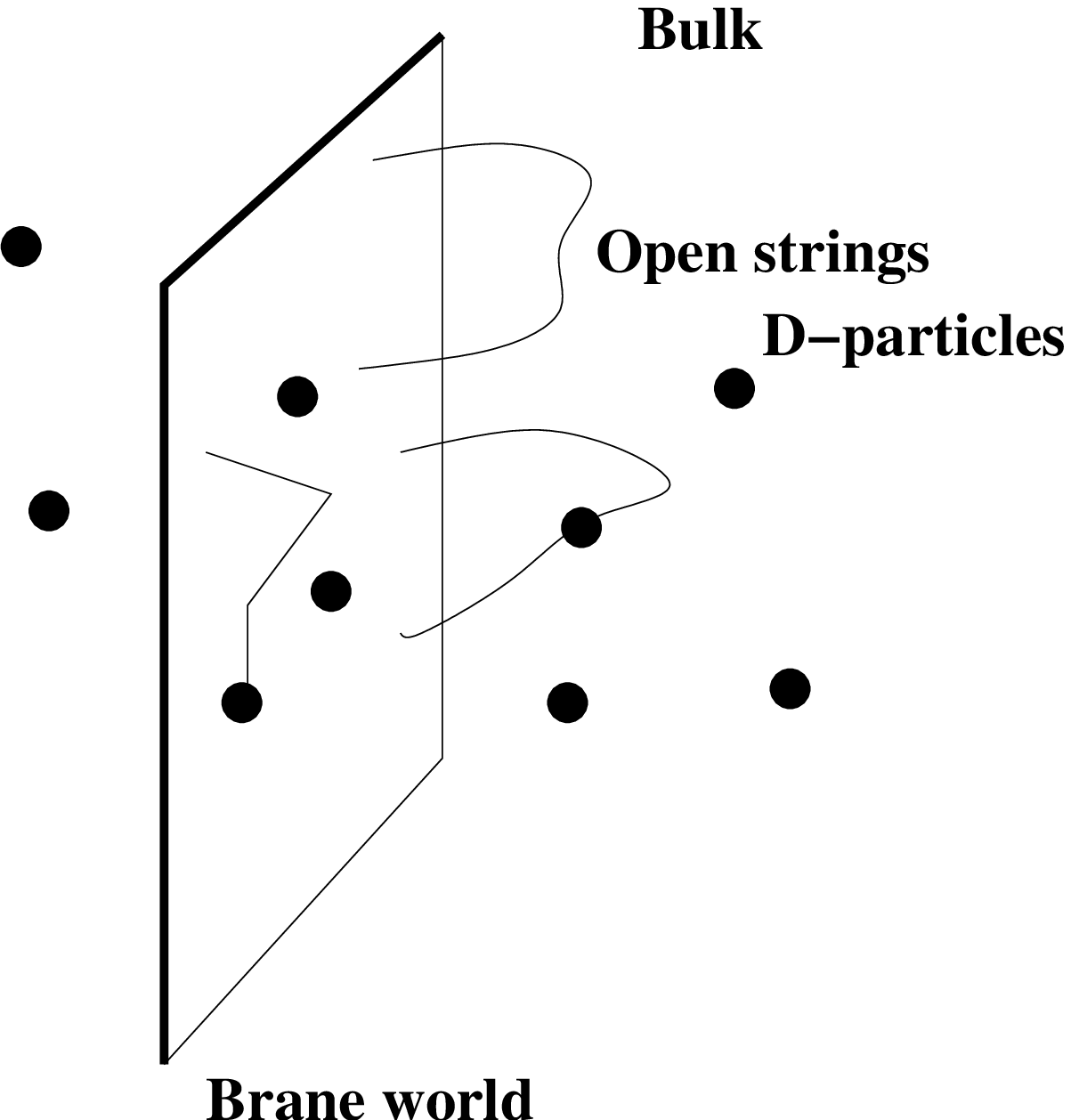}
\end{center}
\caption{\underline{Left}: Recoil of closed string states with
D-particles (space-time defects). \underline{Right}: A
supersymmetric brane world model of D-particle foam. In both cases
the recoil of (massive) D-particle defect causes distortion of
space-time, stochastic metric fluctuations are possible and the
emergent post-recoil string state may differ by flavour and CP
phases.} \label{drecoil}\end{figure}

The discovery of new solitonic structures in superstring theory~\cite{polchinski}
has dramatically changed the understanding of target space structure.
These new non-perturbative objects are known as D-branes and their inclusion leads
to a scattering picture of space-time fluctuations.
A  specific  model  of  stochastic  space-time  foam within the framework of non-critical strings  is  based  on  a
particular kind of gravitational foam~\cite{emn,emw,msdark}, consisting of
``real''   (as   opposed  to   ``virtual'')   space-time  defects   in
higher-dimensional  space   times,  in  accordance   with  the  modern
viewpoint of our  world as a brane hyper-surface  embedded in the bulk
space-time~\cite{polchinski}.   This   model  is  quite   generic  in  some
respects, and we will use it  later to estimate the order of magnitude
of novel CPT violating effects in entangled states of kaons.
In this context one may consider
superstring models of space-time foam, containing a number of point-like
solitonic structures (D-particles)~\cite{emw}, in which motion of D-particles
breaks the target-space supersymmetries.

Heuristically, when low
energy matter given by a closed (or open) string propagating in a $\left(
D+1\right) $-dimensional space-time collides with a very massive D-particle
embedded in this space-time. Since there
are no rigid bodies in general relativity the recoil fluctuations of the
brane and their effectively stochastic back-reaction on space-time cannot be
neglected. On the brane there are closed and open strings propagating. Each
time these strings cross with a D-particle, there is a possibility of being
attached to it (string splitting and capture by the D-particle),
as indicated in Fig. \ref{drecoil}. The entangled state
causes a \emph{back reaction} onto the space-time, which can be calculated
perturbatively using logarithmic conformal field theory formalism~\cite{kmw}.
Now for large Minkowski time $t$, the non-trivial changes from the flat metric produced
from D-particle collisions are
\begin{equation}
g_{0i}\simeq \overline{u}_{i}\equiv \frac{u_{i}}{\varepsilon }\propto \frac{%
\Delta p_{i}}{M_{P}}  \label{recoil}
\end{equation}
where $u_i$ is the velocity and $\Delta p_{i}$ is the momentum transfer during a collision, $\varepsilon ^{-2}$ is identified with $t$ and $M_{P}$
is the Planck mass (actually, to be more precise $M_{P}=M_{s}/g_{s}$, where $%
g_{s}<1$ is the (weak) string coupling, and $M_{s}$ is a string mass scale);
so $g_{0i}$ is constant in space-time but depends on the energy content of
the low energy particle and the Ricci tensor $R_{MN}=0$ where $M$ and $N$
are target space-time indices. Since we are interested in fluctuations of
the metric the indices $i$ will correspond to the pair $M,N$. This master equation will serve as a framework for phenomenological applications.
It must be noted that in this framework the presence of D-particles as well as the splitting/capture processes of string matter illustrated in fig.~\ref{drecoil} lead to a breakdown of the local effective lagrangian formalism. This, in turn, results in ``environmental'' non-Hamiltonian terms in the temporal evolution of the reduced density matrix of the matter string system, which is viewed as an \emph{open quantum system} from the point of view of a low-energy observer on the brane world.

Following
work on gravitational decoherence \cite{emn,msdark}, the target-space
metric state, which is close to being flat, can be represented in this framework
schematically as a density matrix
\begin{equation}
\rho_{\mathrm{grav}}=\int d\,^{5}r\,\,f\left(  r_{\mu}\right)
\left| g\left(  r_{\mu}\right)  \right\rangle \left\langle g\left(
r_{\mu}\,\right)\right|  .\, \label{gravdensity}%
\end{equation}
\bigskip The parameters $r_{\mu}\,\left(  \mu=0,1 \ldots \right)  $ pertain
to appropriate space-time metric deformations and are
\emph{stochastic}, with a Gaussian distribution $\,f\left(  r_{\mu}\,\right)
$
characterized by the averages%
\[
\left\langle r_{\mu}\right\rangle =0,\;\left\langle r_{\mu}r_{\nu
}\right\rangle =\Delta_{\mu}\delta_{\mu\nu}\,.
\]
We will assume that
the fluctuations of the metric felt by two entangled neutral mesons
are independent, and $\Delta_{\mu}\sim O\left(  \frac{E^{2}}%
{M_{P}^{2}}\right)  $, i.e., very small. As matter moves through the
space-time foam in a typical ergodic picture, the effect of time
averaging is assumed to be equivalent to an ensemble average.
For our present discussion we consider a
semi-classical picture for the metric, and therefore $\left|  g\left(
r_{\mu}\right)  \right\rangle $ in (\ref{gravdensity}) is a
coherent state. In the specific model of foam discussed in \cite{msdark}, the
distortion of space-time caused by this scattering can be considered
dominant only along the direction of motion of the matter probe.
Random fluctuations are then considered about an average flat
Minkowski space-time. The result is an effectively  two-dimensional
approximate fluctuating metric describing the main effects~\cite{msdark}

\begin{eqnarray}
g^{\mu\nu}= \left(\begin{array}{cc}
  -(a_1+1)^2 + a_2^2 & -a_3(a_1+1) +a_2(a_4+1) \\
  -a_3(a_1+1) +a_2(a_4+1) & -a_3^2+(a_4+1)^2 \\
\end{array}\right).
\label{flct}
 \end{eqnarray}
The $a_i$ represent the fluctuations and are assumed to be random
variables, satisfying $\langle a_i\rangle =0$ and  $\langle a_i a_j\rangle =
\delta_{ij}\sigma_i$,~$i,j=1,\dots 4$.

Such a (microscopic) model of space-time foam is not of a simple Lindblad
type. The model combines environmental effects, associated with a non-Hamiltonian
evolution, which are associated with the D-particle recoil, with stochastic fluctuations of the
induced space-time metric, which affect the hamiltonian part in such backgrounds.
The resulting effects of such a combined evolution
can be most easily studied~\cite{msdark} by considering the oscillation
probability for, say, two-level scalar systems describing
oscillating neutral mesons, such as Kaons, $K^0 ~\leftrightarrow \overline{K}^0$.
In fact, when curvature effects in space time are ignored, the oscillatory damping effects
due to decoherence have a similar also for fermions, and hence the formulae that follow
also describe neutrino oscillations in such backgrounds.

In the approximation of small fluctuations one
finds the following form for the oscillation probability of the
two-level (scalar) system:
\begin{eqnarray}
\langle e^{i(\omega _{1}-\omega _{2})t}\rangle = \frac{4\tilde{d}^{2}}{(P_{1}P_{2})^{1/2}}\exp \left( \frac{\chi _{1}}{%
\chi _{2}}\right) \exp (i\tilde{b}t)\,, \label{timedep}
\end{eqnarray}%
where $\omega_i,~i=1,2$ are the appropriate energy levels~\cite{msdark}
of the two-level kaon system in the background of the fluctuating
space-time~(\ref{flct}), and

{\small \begin{eqnarray*}
\chi _{1} &=&-4(\tilde{d}^{2}\sigma _{1}+\sigma _{4}k^{4})\tilde{b}%
^{2}t^{2}+2i\tilde{d}^{2}\widetilde{b}^{2}\widetilde{c}k^{2}\sigma
_{1}\sigma _{4}t^{3}, \\
\chi _{2} &=&4\tilde{d}^{2}-2i\tilde{d}^{2}(k^{2}\tilde{c}\sigma _{4}+2%
\tilde{b}\sigma _{1})t+ \nonumber \\
&& \widetilde{b}k^{2}\left( \widetilde{b}k^{2}-2%
\widetilde{d}^{2}\widetilde{c}\right) \sigma _{1}\sigma _{4}, \\
P_{1} &=&4\tilde{d}^{2}+2i\widetilde{d}\widetilde{b}\left( k^{2}-\widetilde{d%
}\right) \sigma _{2}t+\tilde{b}^{2}k^{4}\sigma _{2}\sigma _{3}t^{2}, \\
P_{2} &=&4\tilde{d}^{2}-2i\widetilde{d}^{2}\left(
k^{2}\widetilde{c}\sigma _{4}+2\widetilde{b}\sigma _{1}\right)
t+ \mathcal{O}\left( \sigma ^{2}\right)~,
\end{eqnarray*}}
with {\small \begin{eqnarray*}
\begin{array}{c}
\tilde{b}=\sqrt{k^{2}+m_{1}^{2}}-\sqrt{k^{2}+m_{2}^{2}}, \nonumber \\
\tilde{c}%
=m_{1}^{2}(k^{2}+m_{1}^{2})^{-3/2}-m_{2}^{2}(k^{2}+m_{2}^{2})^{-3/2}, \nonumber \\
\tilde{d}=\sqrt{k^{2}+m_{1}^{2}}\sqrt{k^{2}+m_{2}^{2}}.\nonumber
\end{array}
\end{eqnarray*}}
From this expression one can see~\cite{msdark} that the stochastic
model of space-time foam leads to a modification of oscillation
behavior quite distinct from that of the Lindblad formulation. In
particular,
the transition probability displays a Gaussian time-dependence,
decaying as $e^{-(...)t^2}$, a modification of
the oscillation period, as well as additional power-law fall-off.

On the other hand, within the framework of the D-particle foam, ne can obtain decoherence with damping exponents in the relevant evolution terms of the reduced density matrix of matter, if one goes beyond the Gaussian model of stochastic space time fluctuations.
Indeed, for specific distributions functions $f(x)$ in the recoil velocities of D-particle populations in the foam (c.f. figure~\ref{drecoil}) which are not Gaussian, for instance of Cauchy-Lorentz type~\cite{jean},
\ba
f(x)=\frac{1}{\pi}\frac{\gamma}{x^2+\gamma^2}
\ea
where $\gamma$ is the characteristic scale parameter,
one derives for the respective oscillation probabilities (ignoring the expansion of the Universe for brevity):
\ba
\langle e^{i (\omega_{1}-\omega_{2})t} \rangle\simeq
\exp\Big\{ ikt\Delta-\gamma kt |\Delta|\Big\}~, \quad \Delta=\frac{m_1^2-m_2^2}{2k^2}<<1~, \label{Cauchy}
\ea
where terms of order ${\mathcal O}(m_i/k)^4$ have been disregarded.

{}From this characteristic time-dependence of the decoherence coefficients in various models,
one can obtain bounds
for the fluctuation strength of space-time foam in particle-physics
systems, such as neutral mesons and neutrinos, which we restrict our attention to for the purposes of this presentation. We note at this point that, when
discussing the CPTV effects of foam on entangled states and neutrinos we make use
of this specific model of stochastically fluctuating Gaussian D-particle
foam~\cite{emw,msdark}, in order to demonstrate the effects explicitly
and obtain definite order-of-magnitude estimates~\cite{bms,bmsw}.

Before closing this subsection, a final but important note to the reader is in order.
From the previous discussion it becomes clear that there is no unique characteristic order of magnitude
for the associated decoherence effects, and the situation is highly model dependent.
In some models, the decoherence coefficient scale inversely proportional to some powers of the energy of the probe, while in others they could scale proportional to it.
Hence, generic parametrizations of QG decoherence effects may be misleading.
This should always be kept in mind when one performs phenomenological searches and derives the bounds for QG decoherence effects at various experiments, which will be discussed in
subsequent sections.

\section{Phenomenology of QG Decoherence and CPT Violation with Neutrinos \label{sec:3}}

Neutrino oscillations is one of the most sensitive particle-physics probes of QG-decoherence to date. The presence of QG decoherence effects would affect the profiles of the oscillation probabilities among the various neutrino flavours,
not only by the presence of appropriate {\it damping factors} in front of the
interference oscillatory terms, but also by appropriate modifications of the oscillation period and phase shifts of the oscillation arguments~\cite{lisi,benattineutrino,morgan,icecube,jean,rubbia,poland}.

The first complete phenomenological attempt to fit decoherence models to
atmospheric neutrino data was done in \cite{lisi}, where for simplicity a
two-generation neutrino model with completely positive Lindblad decoherence,
characterised by a single parameter $\gamma$~\cite{benattineutrino}, and leading to exponential
damping with time of the relevant oscillatory terms in the respective
oscillation probabilities, was considered.

Various dependencies on the energy $E$ of the neutrino probes have been
assumed, in a phenomenological fashion, for the Lindblad decoherence
coefficient $\gamma = \gamma_{\rm Lnb} \left(\frac{E}{\mathrm{GeV}}\right)^n~,$
with
$n =0,2,-1$. The sensitivities in the work of \cite{lisi} from atmospheric
neutrinos plus accelerator data at 90\% C.L. can be summarised
by the
following bounds on the parameter $\gamma_{\rm Lnb}$:
\begin{eqnarray}
\gamma_{\rm Lnb}& < & 0.4 \times 10^{-22} ~\mathrm{GeV}~, ~~ n = 0  \nonumber \\
\gamma_{\rm Lnb} & < & 0.9 \times 10^{-27} ~\mathrm{GeV}~, ~~ n = 2  \nonumber \\
\gamma_{\rm Lnb} & < & 0.7 \times 10^{-21} ~\mathrm{GeV}~, ~~ n = -1
\label{early}
\end{eqnarray}
Recently~\cite{lisi}, updated values on these parameters, referred to 95\% C.L.,
have been provided by means of combining solar-neutrino and KamLand data
\begin{eqnarray}
\gamma_{\rm Lnb}& < & 0.67 \times 10^{-24} ~\mathrm{GeV}~, ~~ n = 0  \nonumber \\
\gamma_{\rm Lnb} & < & 0.47 \times 10^{-20} ~\mathrm{GeV}~, ~~ n = 2  \nonumber \\
\gamma_{\rm Lnb}& < & 0.78 \times 10^{-26} ~\mathrm{GeV}~, ~~ n = -1
\label{lisi1}
\end{eqnarray}
It should be remarked that all these bounds should be taken with a grain of
salt, since there is no guarantee that in a theory of quantum gravity $%
\gamma_{\rm Lnb}$ should be the same in all channels, or that the functional
dependence of the decoherence coefficients $\gamma$ on the probe's energy $E$
follows a simple power law. Complicated functional dependencies $\gamma (E)$
might be present.

In the context of our present talk, we shall restrict ourselves to
a brief discussion of a three-generation fit, including QG-decoherence effects, to all presently available data, including LSND results~\cite{lsnd}.
For more details we refer the reader to the literature~\cite{bmsw}.

We assume for simplicity a Lindblad environment, for which the evolution
(\ref{lindevol} applies. However, as we have discussed above, a
combination of both Lindblad environment and stochastically fluctuating
space-time backgrounds (\ref{flct}),
affecting also the Hamiltonian parts of the matter-density-matrix evolution (\ref{lindevol}), could also be in place.
Without a detailed microscopic model, in the three generation case, the precise physical
significance of the decoherence matrix cannot be fully understood.
In \cite{bmsw} we considered the simplified case in which
the matrix $C \equiv (c_{kl})$ (\ref{cmatrix}) is {\it assumed} to be
of the form
 \begin{eqnarray}
  C =   \left(%
\begin{array}{cccccccc}
  c_{11} & 0 & 0 & 0 & 0 & 0 & 0 & 0 \\
  0 & c_{22} & 0 & 0 & 0 & 0 & 0 & 0 \\
  0 & 0 & c_{33} & 0 & 0 & 0 & 0 & c_{38} \\
  0 & 0 & 0 & c_{44} & 0 & 0 & 0 & 0 \\
  0 & 0 & 0 & 0 & c_{55} & 0 & 0 & 0 \\
  0 & 0 & 0 & 0 & 0 & c_{66} & 0 & 0 \\
  0 & 0 & 0 & 0 & 0 & 0 & c_{77} & 0 \\
  0 & 0 & c_{38} & 0 & 0 & 0 & 0 & c_{88} \\
\end{array}%
\right)
\label{cmatrix2}
 \end{eqnarray}
Positivity of the evolution can be guaranteed if and only if
the matrix $C$ is positive
and hence has non-negative eigenvalues.
As discussed in detail in \cite{msdark,bmsw}, such special choices can be realised for models of
the propagation of neutrinos in  models of stochastically
fluctuating environments~\cite{loreti}, where the decoherence term
corresponds to an appropriate double commutator involving operators that entangle with the environment.The quantum-gravity space time
foam may in principle behave as one such stochastic
environment~\cite{msdark}, and it is this point of view
that we critically examined in \cite{bmsw} and review here, in the context of the entirety of the presently available neutrino data.

The simplified form of the $c_{ij}$ matrix given in (\ref{cmatrix2})
implies a matrix  $L_{ij}$ in (\ref{lindevol}) of the form:
 \begin{eqnarray}
    L = \left(%
\begin{array}{cccccccc}
  {\cal D}_{11} & -\Delta_{12} & 0 & 0 & 0 & 0 & 0 & 0 \\
  \Delta_{12} & {\cal D}_{22} & 0 & 0 & 0 & 0 & 0 & 0 \\
  0 & 0 & {\cal D}_{33} & 0 & 0 & 0 & 0 & {\cal D}_{38} \\
  0 & 0 & 0 & {\cal D}_{44} & -\Delta_{13} & 0 & 0 & 0 \\
  0 & 0 & 0 & \Delta_{13} & {\cal D}_{55} & 0 & 0 & 0 \\
  0 & 0 & 0 & 0 & 0 & {\cal D}_{66} & -\Delta_{23} & 0 \\
  0 & 0 & 0 & 0 & 0 & \Delta_{23} & {\cal D}_{77} & 0 \\
  0 & 0 & {\cal D}_{83} & 0 & 0 & 0 & 0 & {\cal D}_{88} \\
\end{array}%
\right)
 \end{eqnarray}
where we have used the notation
$\Delta_{ij}=\frac{m_i^2-m_j^2}{2p}$, and the matrix ${\cal D}_{ij}$
is expressed in terms of the elements of the $C$-matrix:
${\cal D}_{ij}=\sum_{k,l,m} \frac{c_{kl}}{4}\left(-f_{ilm}f_{kmj}
    +f_{kim}f_{lmj}\right)~.$

The probability of a
neutrino of flavor $\nu_{\alpha}$, created at time $t=0$, being converted to a
flavor $\nu_{\beta}$ at a later time t, is calculated in the
Lindblad framework~\cite{lindblad} to be
 \ba
    P_{\nu_{\alpha}\rightarrow \nu_{\beta}}(t)=\Tr [\rho^{\alpha}(t)\rho^{\beta}]
    = \frac{1}{3}+ \frac{1}{2}\sum_{i,j,k}
    e^{\lambda_{k}t}D_{ik}D_{kj}^{-1}
    \rho_{j}^{\alpha}(0)\rho_{i}^{\beta}~.
\label{lindbladprob}
 \ea
where $\lambda_k $ denote the eigenvalues of the Lindblad-decoherence matrix $L_{ij}$.

A detailed analysis~\cite{bmsw} gives the the final expression for the probability as:
 \ba
    \nonumber P_{\nu_{\alpha}\rightarrow \nu_{\beta}}(t)&=& \frac{1}{3}
    + \frac{1}{2}\left\{ \left[ \rho_{1}^{\alpha}
    \rho_{1}^{\beta}
    \cos\left(\frac{|\Omega_{12}|t}{2}\right)
     + \left(\frac{ \Delta {\cal D}_{21}
   \rho_{1}^{\alpha} \rho_{1}^{\beta} }{|\Omega_{12}|}\right)
    \sin\left(\frac{|\Omega_{12}|t}{2}\right)\right]
    e^{({\cal D}_{11}+{\cal D}_{22})\frac{t}{2}}\right.
    \\ \nonumber &+&\left[ \rho_{4}^{\alpha} \rho_{4}^{\beta}
    \cos\left(\frac{|\Omega_{13}|t}{2}\right) + \left(\frac{\Delta {\cal D}_{54}
    \rho_{4}^{\alpha} \rho_{4}^{\beta} }{|\Omega_{13}|}\right)
    \sin\left(\frac{|\Omega_{13}|t}{2}\right) \right]
    e^{({\cal D}_{44}+{\cal D}_{55})\frac{t}{2}}
    \\ \nonumber &+& \left[ \rho_{6}^{\alpha} \rho_{6}^{\beta}
    \cos\left(\frac{|\Omega_{23}|t}{2}\right)
     +\left(\frac{\Delta {\cal D}_{76}
    \rho_{6}^{\alpha} \rho_{6}^{\beta} }{|\Omega_{23}|}\right)
    \sin\left(\frac{|\Omega_{23}|t}{2}\right) \right]
    e^{({\cal D}_{66}+{\cal D}_{77})\frac{t}{2}}
    \\ \nonumber  &+& \left[ \left(\rho_{3}^{\alpha} \rho_{3}^{\beta}+
     \rho_{8}^{\alpha} \rho_{8}^{\beta}\right)
    \cosh\left(\frac{\Omega_{38}t}{2}\right)\right.
    \\ \nonumber &+& \left.  \left(\frac{2{\cal D}_{38}(\rho_{3}^{\alpha}
     \rho_{8}^{\beta} -
    \rho_{8}^{\alpha} \rho_{3}^{\beta}) + \Delta {\cal D}_{83}
    \left(\rho_{3}^{\alpha} \rho_{3}^{\beta} - \rho_{8}^{\alpha}
    \rho_{8}^{\beta}\right)}{\Omega_{38}}\right)
    \sinh\left(\frac{\Omega_{38}t}{2}\right)\right]
    e^{({\cal D}_{33}+{\cal D}_{88})\frac{t}{2}}~. \\
\label{finalnuprob}
 \ea
Above we have used the notation that $\Delta {\cal D}_{ij}={\cal D}_{ii}-{\cal D}_{jj}$.
We have assumed that
 $2|\Delta_{ij}|<|\Delta {\cal D}_{ij}|$ with the consequence that $\Omega_{ij}$,
$ij=12,13,23$ is imaginary. However,
$\Omega_{38}=\sqrt{({\cal D}_{33}-{\cal D}_{88})^2+4{\cal D}_{38}^{2}}$ will be
real. Taking into account mixing, with the up-to-date values of neutrino mass differences and mixing angles, we can proceed into a fit of (\ref{finalnuprob}) to all the presently available data, including the results of the
neutrino sector of the LSND experiment~\cite{lsnd}, and KamLand spectral distortion data~\cite{kamland}.

\begin{figure}[htb]
\centering
\begin{center}
\includegraphics[width=5.0cm]{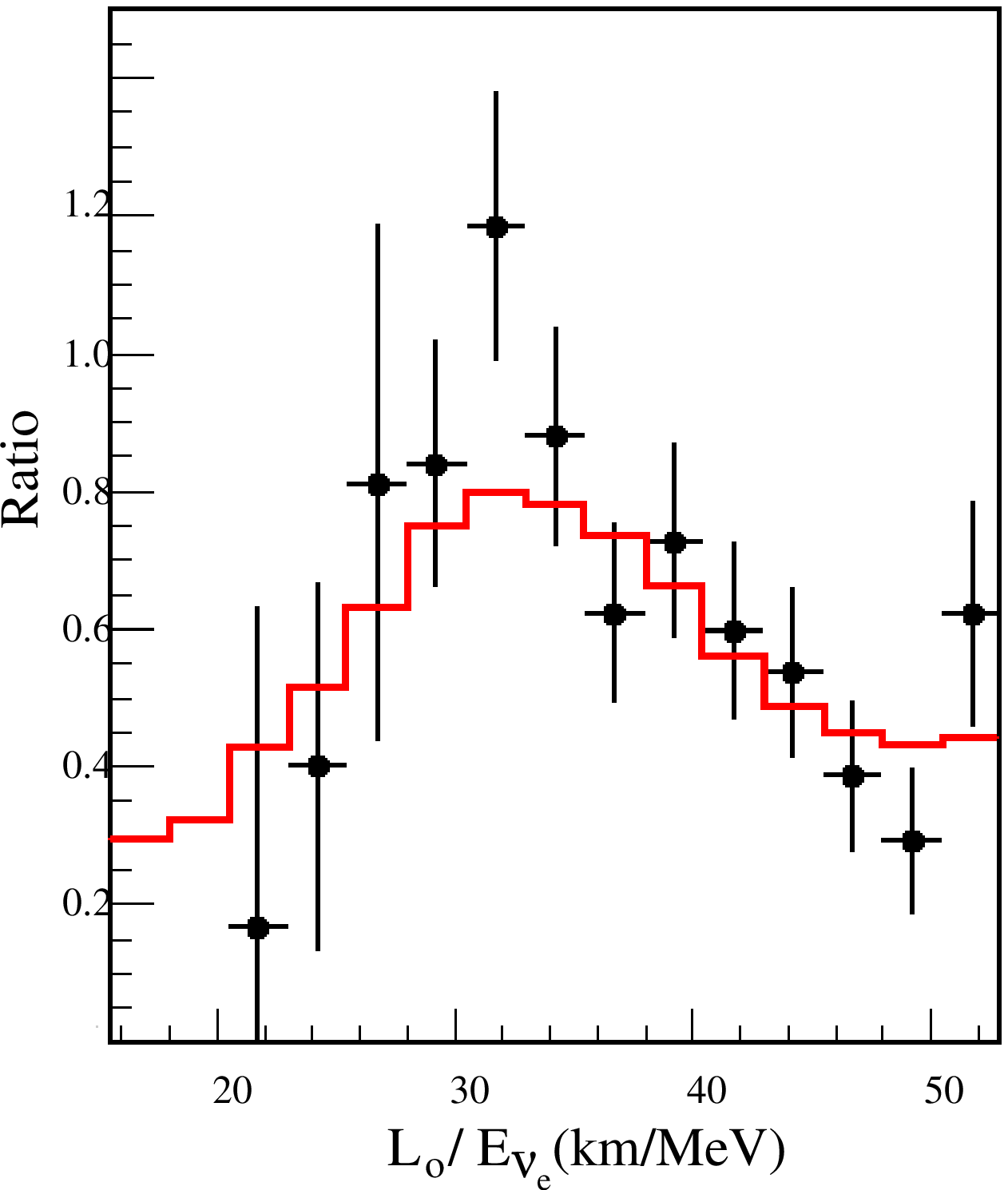}\hspace{.5cm}
\includegraphics[width=6.0cm]{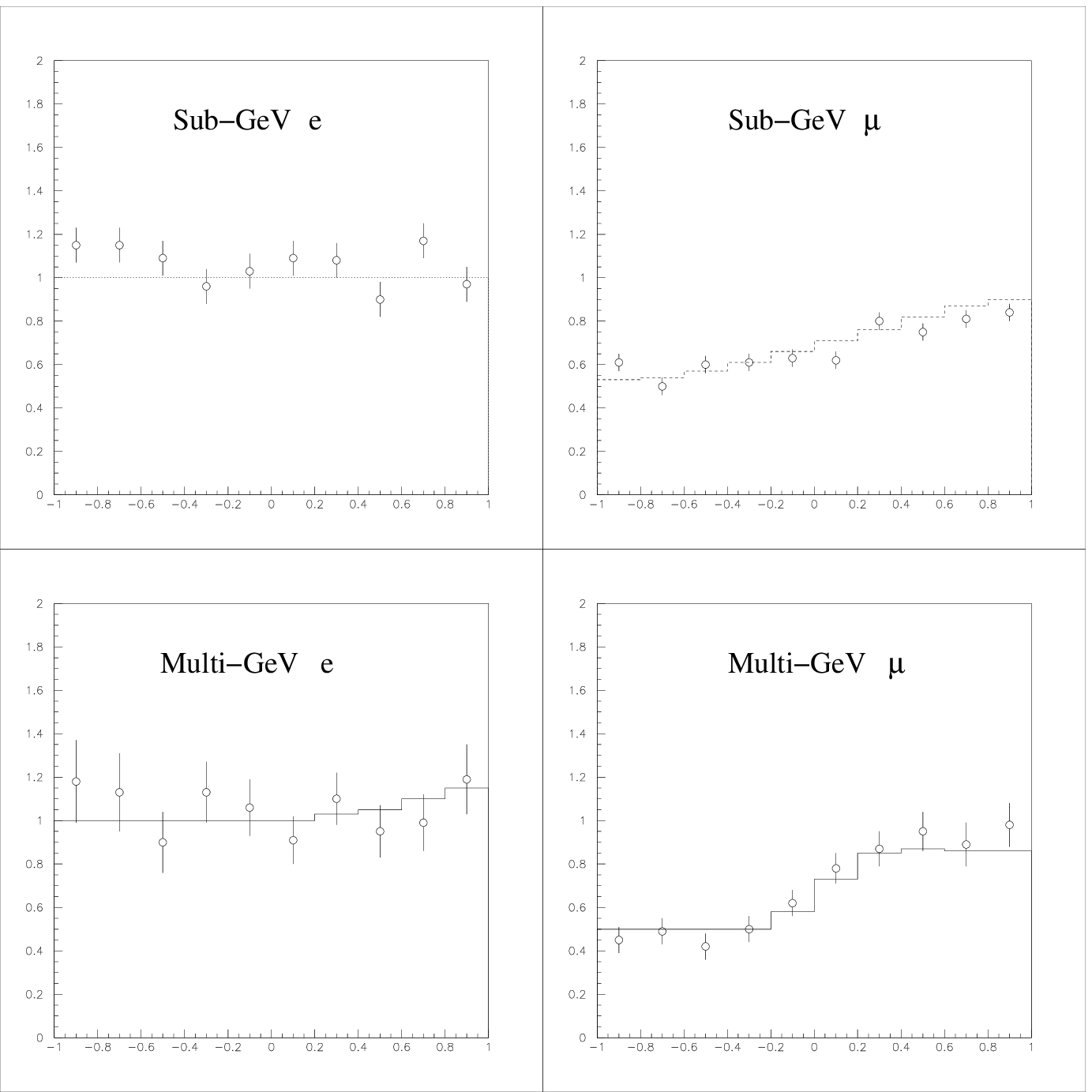}
\end{center}
\caption{\underline{Left}: Ratio of the observed $\overline{\nu}_e$ spectrum to the expectation
versus $L_0/E$ for our decoherence model. The dots correspond to KamLAND data.
\underline{Right}:Decoherence fit. The dots correspond to SK data.}
\label{fig1}
\end{figure}

The results are summarised in Fig.~\ref{fig1},
which demonstrates the agreement (left) of our model
with the KamLand spectral
distortion data~\cite{kamland}, and our best fit (right)
for the Lindblad decoherence model used in ref.~\cite{bmsw},
and in
Table \ref{tab:table1}, where we
present the $\chi^2$ comparison for the model in question
and the standard scenario.

\begin{table}[ht]
\centering
\begin{tabular}{|c|c|c|}
\hline\hline
$\chi^2$ &  decoherence  & standard scenario  \\ [0.5 ex]
\hline\hline
SK  sub-GeV& 38.0  & 38.2 \\ \hline
SK Multi-GeV & 11.7  & 11.2 \\ \hline
Chooz & 4.5 & 4.5  \\\hline
 KamLAND & 16.7 & 16.6 \\\hline
LSND & 0.  & 6.8  \\\hline
TOTAL & 70.9  & 77.3 \\[1ex]
\hline\hline
\end{tabular}
\caption{$\chi^2$ obtained for our model and the one obtained in the standard
scenario for the different experiments calculated with the same program.}
\label{tab:table1}
\end{table}

The best fit has the feature that only {\it some} of the
oscillation terms in the three generation
probability formula have non trivial damping factors,
with their
exponents being {\it independent} of the oscillation
length,
specifically~\cite{bmsw}. If we denote those non trivial
exponents as ${\cal D}\cdot L$, we
obtain from the best fit of \cite{bmsw}:
\ba
{\cal D}=- \frac{\;\;\; 1.3 \cdot 10^{-2}\;\;\;}
{L},
\label{special}
\ea
in units of 1/km with $L=t$ the oscillation length. The
$1/L$-behaviour of ${\cal D}_{11} $, implies, as we mentioned,
oscillation-length independent Lindblad exponents.

In \cite{bmsw} an analysis of the two types of the theoretical models
of space-time foam, discussed in section (\ref{sec:2}), has been performed
in the light of the result of the fit (\ref{special}).
The conclusion was that the Lindblad model of the
stochastically fluctuating media
(\ref{2genprob})
cannot provide the full explanation for the fit, for the following reason:
if the decoherent result of the fit (\ref{special}) was exclusively
due to this model, then the pertinent
decoherent coefficient in that case, for, say, the
KamLand  experiment with an $L \sim 180$~Km,
would be $ |{\cal D}| = \Omega^2 G_N^2 n_0^2 \sim 2.84 \cdot
10^{-21}~{\rm GeV}$ (note that the mixing angle part does not affect the
order of the exponent). Smaller values are found for longer $L$,
such as in atmospheric neutrino experiments or, in future, for high-energy cosmic neutrinos~\cite{morgan,icecube}.
The independence of the
relevant damping exponent from the oscillation length, then, as required
by (\ref{special}) may be understood as follows in this context:
In the spirit of \cite{bm2},
the quantity $G_N n_0 = \xi \frac{\Delta m^2}{E}$,
where $\xi \ll 1$ parametrises the contributions of the foam to the
induced neutrino mass differences, according to our discussion
above. Hence, the damping exponent becomes in this case $ \xi^2
\Omega^2 (\Delta m^2)^2 \cdot L /E^2 $. Thus, for oscillation
lengths $L$ we have
$L^{-1} \sim \Delta m^2/E$, and one is left with  the following
estimate for the dimensionless quantity $\xi^2
\Delta m^2 \Omega^2/E \sim 1.3 \cdot 10^{-2}$. This
implies that the quantity $\Omega^2$ is proportional to the
probe energy $E$. In principle,
this is not an unreasonable result, and it is in
the spirit of \cite{bm2}, since back reaction effects onto
space time, which affect the stochastic fluctuations $\Omega^2$, are
expected to increase with the probe energy $E$. However,
due to the smallness of the quantity $\Delta m^2/E$, for energies
of the order of GeV, and $\Delta m^2 \sim 10^{-3}$ eV$^2$,
we conclude (taking into account that
$\xi \ll 1$) that $\Omega^2$ in this case
would be unrealistically large
for a quantum-gravity effect in the model.

We remark at this point that, in such a model,
we can in principle bound independently the $\Omega$
and $n_0$ parameters by looking at the modifications induced by the
medium in the arguments of the oscillatory functions of the
probability (\ref{2genprob}), that is the period of oscillation.
Unfortunately this is too small to be detected in the above example,
for which $\Delta a_{e\mu} \ll \Delta_{12}$.

The second type of models of stochastically fluctuating space-times
can also be confronted with the data, with similar conclusions.
For instance, in the case of the non-critical string stochastic space-time fluctuation model
(\ref{flct}),(\ref{timedep}), Eq. (\ref{special})
would imply for the pertinent damping exponent
\ba
&& \left(\frac{(m_1^2-m^2_2)^2}{2k^2}
   (9\sigma_1+\sigma_2+\sigma_3+\sigma_4)+
\frac{2V\cos2\theta(m_1^2-m_2^2)}{k}
 (12\sigma_1+2\sigma_2-2\sigma_3)
\right)t^2 \nonumber \\
&& \sim 1.3  \cdot 10^{-2}~.
\ea
Ignoring subleading MSW effects $V$, for simplicity,
and considering oscillation lengths $t=L \sim
\frac{2k}{(m_1^2-m^2_2)}$, we then observe that the independence of
the length $L$ result of the experimental fit, found above, may be
interpreted, in this case, as bounding the stochastic fluctuations
of the metric to $9\sigma_1+\sigma_2+\sigma_3+\sigma_4 \sim
1.3. \cdot 10^{-2}$.
Again, this is too large to be a quantum gravity
effect. Similar conclusions can be reached for the generic stochastically-fluctuating space-time
model (\ref{Pabaverage}).

Such results mean that the $L^2 (=t^2)$ contributions to the
damping due to Gaussian stochastic fluctuations of the metric,
cannot be the exclusive explanation of the fit.

The analysis of \cite{bmsw} also demonstrated that, at least as
far as an order of magnitude of the effect is concerned,
a reasonable explanation of the order of the damping
exponent (\ref{special}), is provided by
Gaussian-type energy fluctuations, due to
ordinary physics effects, leading to decoherence-like damping
of oscillation probabilities~\cite{ohlsson}. The order of these fluctuations,
consistent with
the independence of the damping exponent
on $L$ (irrespective of the power of $L$),
is
\ba
 \frac{\Delta E}{E} \sim 1.6 \cdot 10^{-1}
\ea
if one assumes that this is the principal reason for the
result of the fit. However, the {\it selective} damping implied by the result
of the fit  (\ref{special}), implies that this cannot be the explanation.
The fact that the best fit
model includes terms which are not suppressed at all calls for
a more radical explanation of the fit result, and the issue is
still wide open.

It is interesting, however, that the current neutrino data can
already impose stringent constraints on quantum gravity models, and exclude
some of them from being the
exclusive source of decoherence, as we have discussed above.
Of course, this is not a definite conclusion because one cannot
exclude the possibility of other classes of theoretical models
of quantum gravity, which could escape
these constraints. At present, however, we are not aware of any such
theory.
\begin{figure}[t]
\begin{center}
\includegraphics[width=0.5\textwidth]{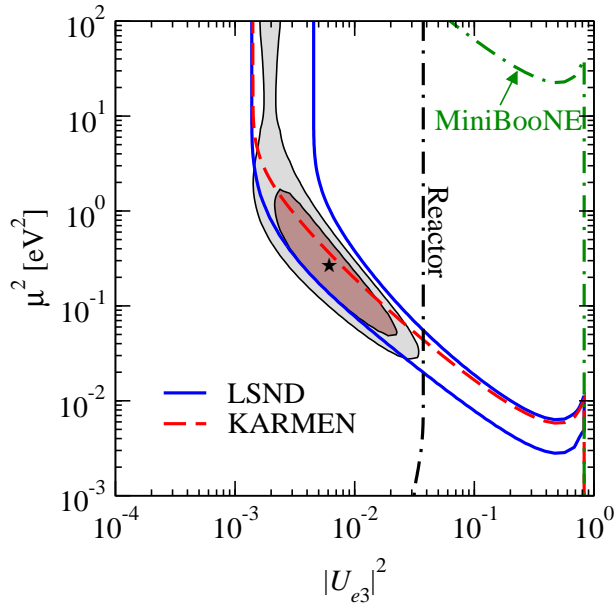}
\end{center}
\caption{Allowed regions in the plane of
    $|U_{e3}|^2$ and the decoherence parameter $\mu^2$ for $r = 4$ in the model of \cite{smirnov}, which reconciles LSND results in the anti-neutrino sector, with those of all the other
    currently available neutrino experiments, inclusive of MiniBoone.
    Shaded regions: global data at 90\% and 99\%~CL. Curves: 99\%~CL
    regions from LSND, KARMEN, MiniBooNE and reactor data (Bugey,
    Chooz, Palo Verde). The star marks the global best fit point. We
    marginalize over $\sin^2\theta_{23}$ taking into account the
    constraint from atmospheric neutrinos.}
    \label{fig:smirnov}
\end{figure}
At this point we would like to remark that the above analysis
took into account the neutrino sector results of the LSND experiment~\cite{lsnd}. In the antineutrino sector, the original indication from that experiment was that there was evidence for much more profound oscillations as compared to the neutrino sector, which, if confirmed, should be interpreted as indicating a direct CPT Violation. There were attempts to interpret such a result as being due to either CPT violating mass differences between neutrinos and antineutrinos, characterising, for instance, non-local theories~\cite{lykken},
or CPT-Violating differences between the strengths of the QG-environmental interactions of neutrinos as compared to those of antineutrinos~\cite{barenboim}, without the need for CPT violating mass differences.

In either case, the order of magnitude of the effects indicated by experiment
was too large for the effect to correspond to realistic QG models, probably pointing towards the need of confirming first the LSND results in the antineutrino sector by future experiments before embarking on radical explanations.

Recently, a purely phenomenological decoherence model of Lindblad type for neutrinos has been suggested in \cite{smirnov}, in an attempt to fit the LSND result in the antineutrino sector with all other available neutrino data.
The model involves \emph{some} damping decoherent coefficients which scale with the neutrino energy as $E^{-4}$. In particular, by assuming masses of the type $m_1 \simeq m_2 \ne m_3$, which is a realistic
pattern for the observed neutrino masses, and considering a decoherence model of Lindblad type, with decoherence coefficients (in a mass-eigenstate basis) of the form:
$\gamma_{12} =0$ and $\gamma_{13}=\gamma_{23} \equiv \gamma$, the authors of \cite{smirnov}
have found the best fit point, which takes into account the LSND together with the other available
neutrino oscillation data, including MiniBoone results in the antineutrino sector, to correspond to
(c.f. figure~\ref{fig:smirnov}):
\begin{equation}
   \gamma = \frac{\mu^2}{E} \left(\frac{40~{\rm MeV}}{E}\right)^3~, \quad \mu^2 = 0.27~{\rm eV}^2,
   \quad |U_{e3}|^2 = 6.1 \times 10^{-3}~, \quad {\rm sin}^2\theta_{23} =0.5~,
\label{lsndsm}
\end{equation}
where $40$ MeV is a typical neutrino energy in the LSND experiment and $U_{ij}$ denote the relevant mixing parameters.

Whether such results can be derived from a realistic model of quantum gravity is not known at present.
We note, however, that as can be seen from the semi-microscopic models of stochastically-fluctuating space-times discussed in section \ref{sec:2}, 
an energy dependence $E^{-4}$ in the decoherence coefficient
is hard to obtain if the stochastically fluctuating space-time is characterised by constant variances. On the other hand, if for some reason, the variances are proportional to appropriate positive powers of $1/k^2$ (with the momentum $k \equiv |\vec k|  \sim E)$ then such dependencies could occur. To understand such issues one needs detailed microscopic models of quantum gravity, with non-trivial back reaction effects of matter onto the space time. At present this is not known.

We would like to close this section with some discussion on the sensitivity of decoherence effects in neutrino oscillations in upcoming facilities, such as CNGS and JPARC~\cite{rubbia}, as well as
astrophysical high-energy cosmic neutrinos~\cite{icecube}.
In \cite{rubbia} we established the limit of sensitivity of CNGS and J-PARC beams, in a
model-independent way, via a simple parametrization of the decoherence effects, discussed in the various theoretical models of section \ref{sec:2}, by
combining in a single model for oscillations between flavours $a,b =1...n$ the various $t=L$ and $t^2=L^2$ dependencies of the decoherence damping factors. The pertinent oscillation probability, averaged over stochastic space-time effects, reads:
\begin{eqnarray}
& \langle P_{\alpha\beta} \rangle= \delta_{\alpha\beta} -  \nonumber \\
& 2 \sum_{a=1}^{n}\sum_{b=1, a<b}^{n}\mathrm{Re}\left( U_{\alpha a}^{*}
U_{\beta a}U_{\alpha b}U_{\beta b}^{*}\right) \left( 1 - \mathrm{cos}%
(2\ell\Delta m_{ab}^{2}) e^{-q_{1}L - q_{2}L^{2}}\right)  \nonumber \\
& -2 \sum_{a=1}^{n} \sum_{b=1, a<b}^{n} \mathrm{Im}\left( U_{\alpha a}^{*}
U_{\beta a}U_{\alpha b}U_{\beta b}^{*}\right) \mathrm{sin}(2\ell\Delta
m_{ab}^{2}) e^{-q_{1}L - q_{2}L^{2}}  \nonumber \\
& ~\mathrm{with} \qquad\ell\equiv\frac{L}{4E}  \label{combinefit}
\end{eqnarray}
where $L \simeq t$ (in units of $c=1$) is the oscillation length. In general
one may parametrize the damping exponents by polynomials in $L$
of any degree, but parametrisations of degrees higher than 2 are not
favoured by the class of quantum-gravity decoherence models considered in
the literature so far~\cite{poland,lisi,msdark}, and reviewed above.

From (\ref{2genprob}), (\ref{Pabaverage}), (\ref{timedep}),  we observe that (\ref{combinefit})
is \emph{oversimplified} in that it ignores possible modifications of the
oscillation period, which do exist in various microscopic models as a result
of the decoherence or stochastic-medium effects. However, for our purposes in the
current article, we note that it is a reasonable assumption that such
modifications to the oscillation period are suppressed as compared with the
ordinary oscillation terms, and as such the dominant, model-independent,
terms appear to be only the exponents of the damping factors.
The parameters $q_{i}, i=1,2$ may be in
general energy dependent, expressing back-reaction effects of the (neutrino)
matter onto the fluctuating space-time. Following earlier treatments and
theoretical quantum-gravity-decoherence models~\cite{lisi,msdark} we
considered the following three cases of generic energy dependence of the
decoherence coefficients $q_{i},i=1,2$:
\begin{equation}\label{qdef}
q_{i},~i=1,2 \quad\propto E^{n}, \quad n=-1, 0, 2
\end{equation}
where the reader should have in mind that in each case the pertinent
decoherence coefficient has the appropriate units, as being a dimensionful
quantity.

The results of the analysis of \cite{rubbia} are summarised in Table~\ref{table_cngs}.
{\
\begin{table}[ht]
\begin{center}
{\scriptsize
\begin{tabular}{|c|c|c|c|}
\hline
Lindblad-type mapping operators &  CNGS & T2K & T2KK \\ \hline
\  &  &  &  \  \\
$\gamma_{0}$\ $[\mathrm{eV}]$\ ;\  ($[\mathrm{GeV}]$)& $2\times10^{-13}$\ ;\
 ($2\times10^{-22}$)& $2.4\times10^{-14}$\ ;\ ($2.4\times10^{-23})$
&
$1.7\times10^{-14}$\ ;\ ($1.7\times10^{-23}$)
\\
\  &  &  &\  \\
$\gamma_{-1}^{2}$\ $[\mathrm{eV^{2}}]$\ ;\  ($[\mathrm{GeV^{2}}]$)&
$9.7\times10^{-4}$\ ;\  ($9.7\times10^{-22}$)&
$3.1\times10^{-5}$\ ;\ ($3.1\times10^{-23}$) &
$6.5\times10^{-5}$\ ;\ ($6.5\times10^{-23}$) \\
\  &  &  &\  \\
$\gamma_{2}$\ $[\mathrm{eV^{-1}}]$\ ;\  ($[\mathrm{GeV^{-1}}]$)&
$4.3\times10^{-35}$\ ;\ ($4.3\times10^{-26}$) &
$1.7\times10^{-32}$\ ;\ ($1.7\times10^{-23}$) &
$3.5\times10^{-33}$\ ;\ ($3.5\times10^{-24}$) \\
\  &  &  &\  \\ \hline
Gravitational MSW (stochastic) effects &  CNGS & T2K & T2KK \\ \hline
\  &  &  &\  \\
$\alpha^{2}$ & $4.3\times10^{-13}~\mathrm{eV}$ & $4.6\times10^{-14}~\mathrm{eV}$
&
$3.5\times10^{-14}~\mathrm{eV}$ \\
\  &  &  &\  \\
$\alpha_{1}^{2}$ & $1.1\times10^{-25}~\mathrm{eV^{2}}$ &
$3.2\times10^{-26}~\mathrm{eV^{2}}$ & $6.7\times10^{-27}~\mathrm{eV^{2}}$\\
\  &  &  &\  \\
$\beta^{2}$ & $3.6\times10^{-24}$ & $5.6\times10^{-23}$ & $1.7\times10^{-23}$\\
\  &  &  &\  \\
$\beta_{2}^{2}$ & $9.8\times10^{-37}~\mathrm{eV}$ &
$4\times10^{-35}~\mathrm{eV}$ & $3.1\times10^{-36}~\mathrm{eV}$ \\
\  &  &  &\  \\
$\beta_{1}^{2}$ & $8.8\times10^{-35}~\mathrm{eV^{-1}}$ &
$3.5\times10^{-32}~\mathrm{eV^{-1}}$ &  $7.2\times10^{-33}~\mathrm{eV^{-1}}$\\
\  &  &  &\  \\ \hline
\end{tabular}
}
\end{center}
\caption{Expected sensitivity limits at CNGS, T2K and T2KK to one parametric
neutrino decoherence
for Lindblad type and gravitational MSW (stochastic metric fluctuation)
like operators. For a detailed explanation of the various decoherence paramaters we refer the reader to Ref.~\cite{rubbia}. We only mention here that they can be expressed in terms of our generic parameters $q_i$ in (\ref{qdef}).}
\label{table_cngs}
\end{table}
}
It is instructive to compare the sensitivity limits presented in
Table~\ref{table_cngs} with those derived from the analysis
of
atmospheric neutrino data~\cite{lisi} obtained at Super-Kamiokande and K2K
experiments. One can transform the limits on the Lindblad type operators
presented in Table~\ref{table_cngs} to the notations of~\cite{lisi}
using the
following transformations:
\begin{eqnarray}
\gamma_{\rm Lnb}& = & \gamma_0[{\rm GeV}], ~~ n = 0 \nonumber
\\
\gamma_{\rm Lnb} & = & \gamma_2[{\rm GeV^{-1}}]\times(\mathrm{GeV}^2) , ~~ n =2
\nonumber \\
\gamma_{\rm Lnb}& = & \gamma_{-1}^2[{\rm GeV^2}]/({\rm GeV}), ~~ n = -1
\label{lisi_transform},
\end{eqnarray}
so that the numbers of Table~\ref{table_cngs} in parentheses can be directly
compared with the bounds~ (\ref{early}) and~ (\ref{lisi1}).
 In particular, the bound obtained in~\cite{lisi} (see for
details~ (\ref{early}) at 95\% C.L. on the
Lindblad type operators with no energy dependence is
close to the sensitivity estimated in our
analysis in case of T2K and T2KK simulations. Although, the CNGS estimation
is about an order of magnitude weaker, one should stress that the current limit
is given at 99\% C.L. under the assumption of the most conservative
level of the uncertainty of the overall neutrino flux at the source. The bound
on the inverse energy dependence given in~\cite{lisi}~(\ref{early}) is close
to the current
CNGS estimates. T2K and T2KK demonstrate an improvement. In spite of the fact that the
Super-Kamiokande data contains neutrino of energies up to $\sim$TeV, the
sensitivity one obtains at CNGS to the energy-squared dependent
decoherence is close, within an order of magnitude, to the bound~(\ref{early})
imposed by atmospheric neutrinos and surpasses T2K and T2KK sensitivity bounds
by $\approx 3$ and $\approx 2$ orders of magnitude respectively. The much less
uncertain systematics of CNGS compared to the atmospheric neutrino data will
make the expected bound more robust as soon as the upcoming data from OPERA will
be analysed.
Moreover, our results are also
competitive with the sensitivity to the same Lindbland operators estimated
in~\cite{morgan} for ANTARES neutrino telescope, which is supposed to
operate at neutrino energies much higher than CNGS and J-PARC
experiments (we note at this stage that, although the sensitivity to the energy squired depending
damping exponent obtained in~\cite{morgan} is very close to our estimations
for CNGS, it is unclear why the authors of~\cite{morgan} claim a
remarkable improvement of this bound relative to the atmospheric
bound~ (\ref{early})).

Assuming that the decoherence phenomena affect all particles in the same
way, which however is by no means certain, one might compare the results of
our analysis with bounds obtained using the neutral kaon system~\cite{cplear}%
. The comparison could be done for the constant (no-energy dependence)
Lindblad decoherence model. The main bound in~\cite{cplear} in such a case
reads $\gamma_{0}\le4.1\times10^{-12}$~eV, thus being about two orders of
magnitude weaker than the sensitivity forecasted in the present paper.

Finally, we compare the estimated sensitivity with the bounds obtained in~%
\cite{lisi} using solar+KamLAND data. In principle, as in the case of the
neutral kaon system, a direct comparison is impossible, since the parameters
investigated here for the $\nu_{\mu}\rightarrow\nu_{\tau}$ channel need not
be the same for the $\nu_{e}\rightarrow\nu_{\mu}$ channel. However, again,
if these parameters are assumed to be roughly of equal size, then one can
see that the estimates of~\cite{lisi} (\ref{lisi1}), which
win essentially over the CNGS,  T2K and T2KK sensitivities only for the case of
inverse energy dependent decoherence, which strongly favours low neutrino
energies.

We would like to close this section by mentioning that strict bounds on decoherence effects
may be expected from observations of very high energy cosmic neutrinos~\cite{icecube}, but the reader should have in mind that this is a highly model dependent statement. Indeed, for models in which the decoherence coefficients are proportional to the energy of the probe,
one might expect very stringent bounds from observations of such cosmic neutrinos,
 since in such a case the decoherence exponents will be of the form ${\cal D} \sim \xi E^{n} \cdot t^n$, with $n$ a positive integer, $t$ the time elapsed from emission till observation and $\xi$ the relevant QG coefficient. In many of these cosmic neutrinos, such times are large, since the distances traveled amount to red-shifts higher than one. Thus, one may expect substantial dumping for very-high-energy neutrinos with energies higher than $10^{18}$ eV.
For models, on the other hand, in which the decoherence damping exponents exhibit
an inverse-power dependence with the energy of the probe, the effects are suppressed for high energy, and lower energy probes may be better for this purpose.

\section{Neutral-Meson Experimental searches for QG Decoherence and CPT Violation \label{sec:4}}

\subsection{Single-Kaon Beam Experiments}

As mentioned in previous sections, QG may induce decoherence
and oscillations $K^0 \leftrightarrow {\overline
K}^0$~\cite{ehns,lopez}, thereby implying a two-level quantum
mechanical system interacting with a QG ``environment''. Adopting
the general assumptions of average energy conservation and monotonic
entropy increase, the simplest model for parametrizing decoherence
(in a rather model-independent way) is the (linear) Lindblad
approach mentioned earlier. Not all entries of a general
decoherence matrix are physical, and in order to isolate the
physically relevant entries one must invoke specific
assumptions, related to the symmetries of the particle system in
question. For the neutral kaon system, such an extra assumptions
is that the QG medium respects
the $\Delta S=\Delta Q$ rule.
In such a case, the modified Lindblad evolution equation
(\ref{evoleq}) for
the respective density matrices of neutral kaon matter can be
parametrized as follows~\cite{ehns}:
$$\partial_t \rho = i[\rho, H] + \delta\H \rho~,$$
where {\small $$H_{\alpha\beta}=\left( \begin{array}{cccc}  - \Gamma
& -\coeff{1}{2}\delta \Gamma
& -{\rm Im} \Gamma _{12} & -{\rm Re}\Gamma _{12} \\
 - \coeff{1}{2}\delta \Gamma
  & -\Gamma & - 2{\rm Re}M_{12}&  -2{\rm Im} M_{12} \\
 - {\rm Im} \Gamma_{12} &  2{\rm Re}M_{12} & -\Gamma & -\delta M    \\
 -{\rm Re}\Gamma _{12} & -2{\rm Im} M_{12} & \delta M   & -\Gamma
\end{array}\right) $$} and
$$ {\delta\H}_{\alpha\beta} =\left( \begin{array}{cccc}
 0  &  0 & 0 & 0 \\
 0  &  0 & 0 & 0 \\
 0  &  0 & -2\alpha  & -2\beta \\
 0  &  0 & -2\beta & -2\gamma \end{array}\right)~.$$
Positivity of $\rho$ requires: $\alpha, \gamma  > 0,\quad
\alpha\gamma>\beta^2$. Notice that $\alpha,\beta,\gamma$ violate
{\it both}  CPT, due to their decohering nature~\cite{wald}, and CP
symmetry, as they do not commute with the CP operator
$\widehat{CP}$~\cite{lopez}: $\widehat{CP} = \sigma_3 \cos\theta +
\sigma_2 \sin\theta$,$~~~~~[\delta\H_{\alpha\beta}, \widehat{CP} ]
\ne 0$.

An important remark is now in order. As pointed out in
\cite{benatti}, although the above parametrization is sufficient for
a single-kaon state to have a positive definite density matrix (and
hence probabilities) this is \emph{not} true when one considers
the evolution of entangled kaon states ($\phi$-factories).
In this latter case,
complete positivity is guaranteed only if
the further conditions
\begin{equation}\label{cons}
\alpha = \gamma~ {\rm and} ~\beta = 0
\end{equation}
are imposed. When incorporating entangled states, one should either
consider possible new effects (such as the $\omega$-effect
considered below) or apply the constraints (\ref{cons}) also to
single kaon states~\cite{benatti}. This is not necessarily the case
when other non-entangled particle states, such as neutrinos, are
considered, in which case the $\alpha,\beta,\gamma$ parametrization
of decoherence may be applied. Experimentally the complete
positivity hypothesis can be tested explicitly. In what follows, as
far as single-kaon states are concerned, we keep the
$\alpha,\beta,\gamma$ parametrization, and give the available
experimental bounds for them, but we always have in mind the
constraint (\ref{cons}) when referring to entangled kaon states in a
$\phi$-factory.

As already mentioned, when testing CPT symmetry with neutral kaons
one should be careful to distinguish two types of CPTV: {\bf (i)} CPTV
within Quantum Mechanics~\cite{fide}, leading to possible
differences between particle-antiparticle masses and widths: $\delta
m= m_{K^0} - m_{{\overline K}^0}$, $\delta \Gamma = \Gamma_{K^0}-
\Gamma_{{\overline K}^0} $. This type of CPTV could be,
for instance,
due to (spontaneous) Lorentz violation~\cite{sme}. In that
case the CPT operator is well-defined as a quantum mechanical
operator, but does not commute with the Hamiltonian of the system.
This, in turn, may lead to mass and width differences between particles
and antiparticles, among other effects. {\bf (ii)} CPTV through
decoherence~\cite{ehns,poland} via the parameters
$\alpha,\beta,\gamma$ (entanglement with the QG ``environment'',
leading to modified evolution for $\rho$ and $\$ \ne S~S^\dagger $).
In the latter case the CPT operator may not be well-defined, which
implies novel effects when one uses entangled states of kaons, as we
shall discuss in the next subsection.

\begin{figure}[htb]
\begin{center}
  \includegraphics[width=0.3\textwidth]{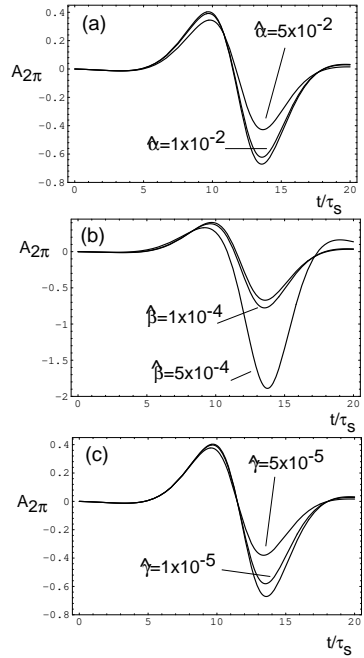}
 \end{center}
\caption{Neutral kaon decay asymmetries $A_{2\pi}$~\cite{lopez}, as a typical example indicating the effects of QG-induced decoherence.}
\label{AT}
\end{figure}

\begin{table}[thb]
\begin{center}
\begin{tabular}{lcc}
\underline{Process}&QMV&QM\\
$A_{2\pi}$&$\not=$&$\not=$\\
$A_{3\pi}$&$\not=$&$\not=$\\
$A_{\rm T}$&$\not=$&$=$\\
$A_{\rm CPT}$&$=$&$\not=$\\
$A_{\Delta m}$&$\not=$&$=$\\
$\zeta$&$\not=$&$=$
\end{tabular}
\caption{Qualitative comparison of predictions for various
observables in CPT-violating theories beyond (QMV) and within (QM)
quantum mechanics. Predictions either differ ($\not=$) or agree
($=$) with the results obtained in conventional quantum-mechanical
CP violation. Note that these frameworks can be qualitatively
distinguished via their predictions for $A_{\rm T}$, $A_{\rm CPT}$,
$A_{\Delta m}$, and $\zeta$.} \label{Table2}
\end{center}
\hrule
\end{table}

The important point to notice is that the two types of CPTV can be
{\it disentangled experimentally}~\cite{lopez}. The relevant
observables are defined as $ \VEV{O_i}= {\rm Tr}\,[O_i\rho] $. For
neutral kaons, one looks at decay asymmetries for $K^0, {\overline
K}^0$, defined as:
$$A (t) = \frac{
    R({\bar K}^0_{t=0} \rightarrow
{\bar f} ) -
    R(K^0_{t=0} \rightarrow
f ) } { R({\bar K}^0_{t=0} \rightarrow {\bar f} ) +
    R(K^0_{t=0} \rightarrow
f ) }~,$$ where $R(K^0\rightarrow f) \equiv \Tr[O_{f}\rho (t)]=$
denotes the decay rate into the final state $f$ (starting from a
pure $ K^0$ state at $t=0$).

In the case of neutral kaons, one may consider the following set of
asymmetries: (i) {\it identical final states}: $f={\bar f} = 2\pi $:
$A_{2\pi}~,~A_{3\pi}$, (ii) {\it semileptonic} : $A_T$ (final states
$f=\pi^+l^-\bar\nu\ \not=\ \bar f=\pi^-l^+\nu$), $A_{CPT}$
(${\overline f}=\pi^+l^-\bar\nu ,~ f=\pi^-l^+\nu$), $A_{\Delta m}$.
Typically, for instance when final states are $2\pi$, one has  a
time evolution of the decay rate $R_{2\pi}$: $ R_{2\pi}(t)=c_S\,
e^{-\Gamma_S t}+c_L\, e^{-\Gamma_L t} + 2c_I\, e^{-\Gamma
t}\cos(\Delta mt-\phi)$, where $S$=short-lived, $L$=long-lived,
$I$=interference term, $\Delta m = m_L - m_S$, $\Delta \Gamma =
\Gamma_S - \Gamma_L$, $\Gamma =\frac{1}{2}(\Gamma_S + \Gamma_L)$.
One may define the {\it decoherence parameter}
$\zeta=1-{c_I\over\sqrt{c_Sc_L}}$, as a (phenomenological) measure
of quantum decoherence induced in the system~\cite{fide}. For larger
sensitivities one can look at this parameter in the presence of a
regenerator~\cite{lopez}. In our decoherence scenario, $\zeta$
corresponds to a particular combination of the decoherence
parameters~\cite{lopez}:
$$ \zeta \to \frac{\widehat \gamma}{2|\epsilon ^2|} -
2\frac{{\widehat \beta}}{|\epsilon|}{\rm sin} \phi~,$$ with the
notation $\widehat{\gamma} =\gamma/\Delta \Gamma $, \emph{etc}.
Hence, ignoring the constraint (\ref{cons}), the best bounds on
$\beta$, or -turning the logic around- the most sensitive tests of
complete positivity in kaons, can be placed by implementing a
regenerator~\cite{lopez}.

The experimental tests (decay asymmetries) that can be performed in
order to disentangle decoherence from quantum-mechanical CPT
violating effects are summarized in Table \ref{Table2}. In Figure
\ref{AT} we give a typical profile of a
decay asymmetries~\cite{lopez}, from where bounds on QG decohering
parameters can be extracted. At present there are experimental bounds
available from CPLEAR measurements~\cite{cplear} $\alpha
< 4.0 \times 10^{-17} ~{\rm GeV}~, ~|\beta | < 2.3. \times 10^{-19}
~{\rm GeV}~, ~\gamma < 3.7 \times 10^{-21} ~{\rm GeV} $, which are
not much different from theoretically expected values in some
optimistic scenarios~\cite{lopez} $\alpha~,\beta~,\gamma = O(\xi
\frac{E^2}{M_{P}})$.

Recently, the experiment KLOE at DA$\Phi$NE updated these limits by
measuring for the first time the $\gamma$ decoherence parameter for
entangled kaon states~\cite{adidomenico}, as well as the (naive)
decoherence parameter $\zeta$ (to be specific, the KLOE
Collaboration has presented measurements for two $\zeta$ parameters,
one, $\zeta_{LS}$, pertaining to an expansion in terms of $K_L,K_S$
states, and the other, $\zeta_{0\bar 0}$, for an expansion in terms
of $K^0,\overline K^0$  states). We remind the reader once more
that, under the assumption of complete positivity for entangled
meson states~\cite{benatti}, theoretically there is only one
parameter to parametrize Lindblad decoherence, since $\alpha =
\gamma$, $\beta = 0$. In fact, the KLOE experiment has the greatest
sensitivity to this parameter $\gamma$. The latest KLOE
measurement for $\gamma$, as reported
by A. DiDomenico in this conference~\cite{adidomenico},
yields $\gamma_{\rm
KLOE} = (0.7^{+1.2}_{-1.2} \pm 0.3) \times 10^{-21}~{\rm GeV}$, i.e.
$\gamma < 7 \times 10^{-22}~{\rm GeV}$, competitive with the
corresponding CPLEAR bound~\cite{cplear} discussed above. It is
expected that this bound could be improved by an order of magnitude
in upgraded facilities, such as KLOE-2 at
DA$\Phi$NE-2~\cite{adidomenico}, where one expects $\gamma_{\rm
upgrade} \to \pm 0.2 \times 10^{-21} ~{\rm GeV}$.

The reader should also bear in mind that the Lindblad linear
decoherence is not the only possibility for a parametrization of QG
effects, see for instance the stochastically fluctuating space-time
metric approach discussed in Section 3.1 above. Thus, direct tests of
the complete positivity hypothesis in entangled states, and hence
the theoretical framework {\it per se}, should be performed by
independent measurements of all the three decoherence parameters
$\alpha,\beta,\gamma$; as far as we understand, such data
are currently available in kaon factories, but not yet analyzed in
detail~\cite{adidomenico}.

\subsection{ Entangled Neutral Meson States, Modified EPR Correlations
and CPT Violation}

\subsubsection{The $\omega$-Effect}

We now come to a description of an entirely novel effect~\cite{bmp}
of CPTV due to the ill-defined nature of the CPT
operator, which is \emph{exclusive} to neutral-meson factories, for
reasons explained below. The effects are qualitatively similar
for kaon and $B$-meson factories~\cite{bomega}, with the
important observation that in kaon factories there is a particularly
good channel, that of both correlated kaons decaying to $\pi^+\pi^-$.
In that channel the sensitivity of the effect increases because
the complex parameter $\omega$, parametrizing the relevant
EPR modifications~\cite{bmp}, appears in the particular
combination $|\omega|/|\eta_{+-}|$, with
$|\eta_{+-}| \sim 10^{-3}$. In the case of  $B$-meson factories
one should focus instead on the ``same-sign'' di-lepton
channel~\cite{bomega}, where high statistics is expected.

We commence
our discussion by briefly reminding the reader of
EPR particle correlations.
The EPR effect was originally proposed as a {\it paradox}, testing the
foundations of Quantum Theory. There was the question whether
quantum correlations between spatially separated events implied
instant transport of information that would contradict special relativity.
It was eventually realized that no super-luminal propagation was
actually involved in the EPR phenomenon, and thus there was no
conflict with relativity.

\begin{figure}[ht]
\begin{center}
  \includegraphics[width=0.3\textwidth]{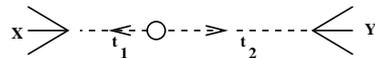}
\end{center}
\caption{Schematic representation of the decay of a $\phi$-meson at
rest (for definiteness) into pairs of entangled neutral kaons, which
eventually decay on the two sides of the detector.}
\label{epr}\end{figure}

The EPR effect has been confirmed experimentally, e.g., in meson
factories: (i) a pair of particles can be created in a definite
quantum state, (ii) move apart and, (iii) eventually decay when they
are widely (spatially) separated (see Fig.~\ref{epr} for a schematic
representation of an EPR effect in a meson factory). Upon making a
measurement on one side of the detector and identifying the decay products,
we \emph{infer} the type of products appearing on the
other side; this is essentially the EPR correlation phenomenon.
It does \emph{not} involve any \emph{simultaneous measurement} on
both sides, and hence there is no contradiction with special
relativity. As emphasized by Lipkin~\cite{lipkin}, the EPR
correlations between different decay modes should be taken into
account when interpreting any experiment.

In the case of $\phi$ factories
it was \emph{claimed }~\cite{dunietz} that
due to EPR correlations,  \emph{irrespective} of CP, and CPT
violation, the \emph{final state} in $\phi$ decays: $e^+ e^-
\Rightarrow \phi \Rightarrow K_S K_L $ always contains $K_LK_S$
products.
This is a direct consequence
of imposing the requirement of {\it Bose statistics}
on the state $K^0 {\overline K}^0$ (to which the $\phi$ decays);
this, in turn, implies that the physical neutral meson-antimeson
state must be {\it symmetric} under C${\cal P}$, with C the charge
conjugation and ${\cal P}$ the operator that permutes the spatial
coordinates. Assuming {\it conservation} of angular momentum, and a
proper existence of the {\it antiparticle state} (denoted by a bar),
one observes that: for $K^0{\overline K}^0$ states which are
C-conjugates with C$=(-1)^\ell$ (with $\ell$ the angular momentum
quantum number), the system has to be an eigenstate of the
permutation operator ${\cal P}$ with eigenvalue $(-1)^\ell$. Thus,
for $\ell =1$: C$=-$ $\rightarrow {\cal P}=-$.  Bose statistics
ensures that for $\ell = 1$ the state of two \emph{identical} bosons
is \emph{forbidden}. Hence, the  initial entangled state:

{\begin{eqnarray*} &&|i> = \frac{1}{\sqrt{2}}\left(|K^0({\vec
k}),{\overline K}^0(-{\vec k})>
- |{\overline K}^0({\vec k}),{K}^0(-{\vec k})>\right)  \nonumber \\
&& = {\cal N} \left(|K_S({\vec k}),K_L(-{\vec k})> - |K_L({\vec
k}),K_S(-{\vec k})> \right)\nonumber
\end{eqnarray*}}with the normalization factor ${\cal N}=\frac{\sqrt{(1
+ |\epsilon_1|^2) (1 + |\epsilon_2|^2
)}}{\sqrt{2}(1-\epsilon_1\epsilon_2)} \simeq \frac{1 +
|\epsilon^2|}{\sqrt{2}(1 - \epsilon^2)}$, and
$K_S=\frac{1}{\sqrt{1 + |\epsilon_1^2|}}\left(|K_+> + \epsilon_1
|K_->\right)$, $K_L=\frac{1}{\sqrt{1 + |\epsilon_2^2|}}\left(|K_->
+ \epsilon_2 |K_+>\right)$, where $\epsilon_1, \epsilon_2$ are
complex parameters, such that $\epsilon \equiv \epsilon_1 +
\epsilon_2$ denotes the CP- \& T-violating parameter, whilst $\delta
\equiv \epsilon_1 - \epsilon_2$  parametrizes the CPT \& CP violation
within quantum mechanics~\cite{fide}, as discussed previously.
The $K^0 \leftrightarrow {\overline K}^0$ or $K_S
\leftrightarrow K_L$ correlations are apparent after evolution, at
any time $t > 0$ (with $t=0$ taken as the moment of the $\phi$ decay).

In the above considerations there is an implicit assumption,
which was noted in \cite{bmp}. The above arguments are valid
independently of CPTV, provided such violation occurs
within quantum mechanics, e.g., due to spontaneous Lorentz
violation, where the CPT operator is well defined.

If, however, CPT is \emph{intrinsically} violated, due, for instance,
to decoherence scenarios in space-time foam, then
the factorizability property of the
super-scattering matrix \$ breaks down, \$ $\ne SS^\dagger $, and
the generator of CPT is not well defined~\cite{wald}. Thus, the
concept of an ``antiparticle'' may be \emph{modified} perturbatively! The
perturbative modification of the properties of the antiparticle is
important, since the antiparticle state is a physical state which
exists, despite the ill-definition of the CPT operator. However, the
antiparticle Hilbert space will have components that are
\emph{independent} of the particle Hilbert
space.

In such a case,
the neutral mesons $K^0$ and ${\overline K}^0$ should \emph{no
longer} be treated as \emph{indistinguishable particles}. As a
consequence~\cite{bmp}, the initial entangled state in $\phi$
factories $|i>$, after the $\phi$-meson decay, will acquire a component
with opposite permutation (${\cal P}$) symmetry:

{ \begin{eqnarray}\label{initialomega} |i> &=& \frac{1}{\sqrt{2}}\left(|K_0({\vec
k}),{\overline K}_0(-{\vec k})>
- |{\overline K}_0({\vec k}),K_0(-{\vec k})> \right)\nonumber \\
&+&  \frac{\omega}{2} \left(|K_0({\vec k}), {\overline K}_0(-{\vec k})> + |{\overline K}_0({\vec
k}),K_0(-{\vec k})> \right)  \bigg]  \nonumber \\
& = & \bigg[ {\cal N} \left(|K_S({\vec
k}),K_L(-{\vec k})>
- |K_L({\vec k}),K_S(-{\vec k})> \right)\nonumber \\
&+&  \omega \left(|K_S({\vec k}), K_S(-{\vec k})> - |K_L({\vec
k}),K_L(-{\vec k})> \right)  \bigg]~,
\end{eqnarray}}where ${\cal N}$ is an appropriate normalization factor,
and $\omega = |\omega |e^{i\Omega}$ is a complex parameter,
parametrizing the intrinsic CPTV modifications of the EPR
correlations. Notice that, as a result of the $\omega$-terms, there
exist, in the two-kaon state,
$K_SK_S$ or $K_LK_L$ combinations,
which
entail important effects to the various decay channels. Due to this
effect, termed the $\omega$-effect by the authors of \cite{bmp},
there is \emph{contamination} of ${\cal P}$(odd) state with ${\cal P}$({\rm even})
terms. The $\omega$-parameter controls the amount of contamination
of the final ${\cal P}$(odd) state by the ``wrong'' (${\cal P}$(even)) symmetry state.

Later in this section  we will present a microscopic model where such a
situation is realized explicitly. Specifically,
an $\omega$-like effect appears due to the evolution in the space-time
foam, and the corresponding parameter turns out to be
purely imaginary and time-dependent~\cite{bms}.

\subsubsection{$\omega$-Effect Observables in $\phi$-factories}

To construct the appropriate observable for the possible detection
of the $\omega$-effect, we consider the $\phi$-decay amplitude
depicted in Fig.~\ref{epr}, where one of the kaon products decays to
the  final state $X$ at $t_1$ and the other to the final state $Y$
at time $t_2$. We take $t=0$ as the moment of the $\phi$-meson
decay.

The relevant amplitudes read:
\begin{eqnarray*}
A(X,Y) = \langle X|K_S\rangle \langle Y|K_S \rangle \, {\cal N}
\,\left( A_1  +  A_2 \right)~, \nonumber
\end{eqnarray*}
with \begin{eqnarray*}
 A_1  &=& e^{-i(\lambda_L+\lambda_S)t/2}
[\eta_X  e^{-i \Delta\lambda \Delta t/2}
-\eta_Y  e^{i \Delta\lambda \Delta t/2}]\nonumber \\
A_2  &=&  \omega [ e^{-i \lambda_S t} - \eta_X \eta_Y e^{-i
\lambda_L t}] \nonumber
\end{eqnarray*}
denoting the CPT-allowed and CPT-violating parameters respectively,
and $\eta_X = \langle X|K_L\rangle/\langle X|K_S\rangle$ and $\eta_Y
=\langle Y|K_L\rangle/\langle Y|K_S\rangle$. In the above formulae, $t$
is the sum of the decay times $t_1, t_2$ and $\Delta t $ is their
difference (assumed positive).

\begin{figure}[htb]
\begin{center}
  \includegraphics[width=0.3\textwidth]{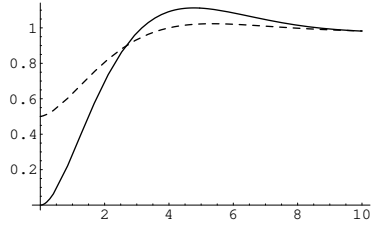}
\end{center}
\caption{A characteristic case of the intensity $I(\Delta t)$, with
$|\omega|=0$ (solid line)  vs  $I(\Delta t)$ (dashed line) with
$|\omega|=|\eta_{+-}|$, $\Omega = \phi_{+-} - 0.16\pi$, for
definiteness~\cite{bmp}.} \label{intensomega}\end{figure}

The ``intensity'' $I(\Delta t)$ is the
desired \emph{observable} for a detection of the $\omega$-effect,
\begin{eqnarray} \label{omegevoldet} I (\Delta t) \equiv \frac{1}{2} \int_{\Delta
t}^\infty dt\, |A(X,Y)|^2~.
\end{eqnarray}
depending only on $\Delta t$.

Its time profile reads~\cite{bmp}:
\begin{eqnarray}\label{omegevoldet2} &&
I (\Delta t) \equiv \frac{1}{2} \int_{|\Delta t|}^\infty dt\,
|A(\pi^+\pi^-,\pi^+\pi^-)|^2  = \nonumber
\\ && |\langle\pi^{+}\pi^{-}|K_S\rangle|^4 |{\cal N}|^2 |\eta_{+-}|^2
\bigg[ I_1  + I_2  +  I_{12} \bigg]~,
\end{eqnarray}
where
\begin{eqnarray}\label{omegevoldet3} && I_1 (\Delta t) =
\frac{e^{-\Gamma_S \Delta t} + e^{-\Gamma_L \Delta t} - 2
e^{-(\Gamma_S+\Gamma_L) \Delta t/2} \cos(\Delta m \Delta t)}
{\Gamma_L+\Gamma_S}
\nonumber \\
&& I_2 (\Delta t) =  \frac{|\omega|^2 }{|\eta_{+-}|^2}
\frac{e^{-\Gamma_S \Delta t} }{2 \Gamma_S}
\nonumber \\
&& I_{12} (\Delta t) = - \frac{4}{4 (\Delta m)^2 + (3 \Gamma_S +
\Gamma_L)^2}  \frac{|\omega|}{|\eta_{+-}|} \times
\nonumber \\
&&\bigg[ 2 \Delta m \bigg( e^{-\Gamma_S \Delta t} \sin(\phi_{+-}-
\Omega) - \nonumber \\
&&  e^{-(\Gamma_S+\Gamma_L) \Delta t/2} \sin(\phi_{+-}- \Omega
+\Delta m \Delta t)\bigg)
\nonumber \\
&&  - (3 \Gamma_S + \Gamma_L) \bigg(e^{-\Gamma_S \Delta t}
\cos(\phi_{+-}- \Omega) - \nonumber \\
&& e^{-(\Gamma_S+\Gamma_L) \Delta t/2} \cos(\phi_{+-}- \Omega
+\Delta m \Delta t)\bigg)\bigg]~,
\end{eqnarray}
with $\Delta m = m_S - m_L$ and $\eta_{+-}= |\eta_{+-}|
e^{i\phi_{+-}}$ in the usual notation~\cite{fide}.

A typical case for the relevant intensities, indicating clearly the
novel CPTV $\omega$-effects, is depicted in Fig.~\ref{intensomega}.

As announced, the novel $\omega$-effect appears in
the combination $\frac{|\omega|}{|\eta_{+-}|}$, thereby implying
that the decay channel to $\pi^+\pi^-$ is particularly sensitive to
the $\omega$ effect~\cite{bmp}, due to the enhancement by
$1/|\eta_{+-}| \sim 10^{3}$, implying sensitivities up to
$|\omega|\sim 10^{-6}$ in $\phi$ factories. The physical reason for
this enhancement is that $\omega$ enters through $K_SK_S$ as opposed to
$K_LK_S$ terms, and the $K_L \to \pi^+\pi^-$ decay is CP-violating.

\subsubsection{B-Meson Factories and the $\omega$-effect }

\begin{figure}
\begin{center}
  \includegraphics[width=6cm]{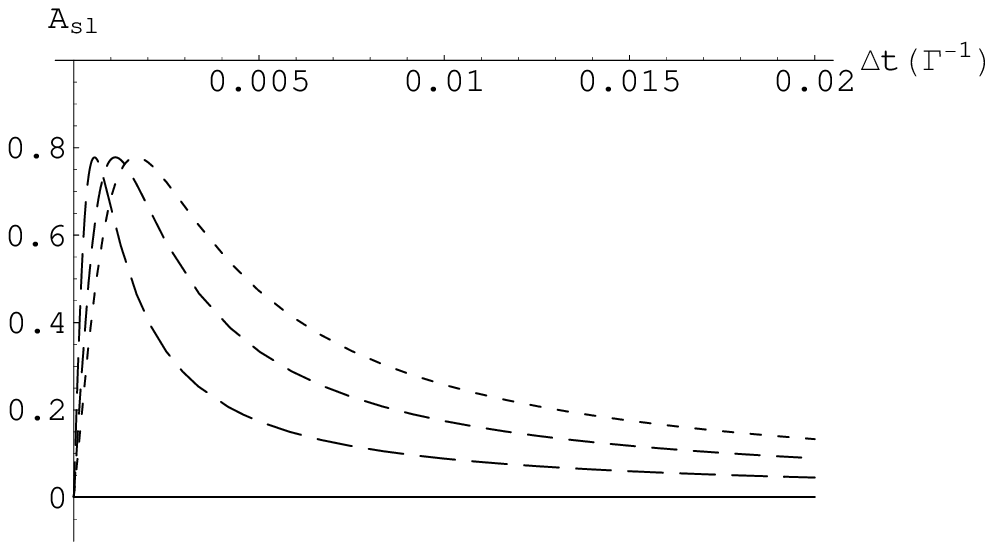}\hfill \includegraphics[width=6cm]{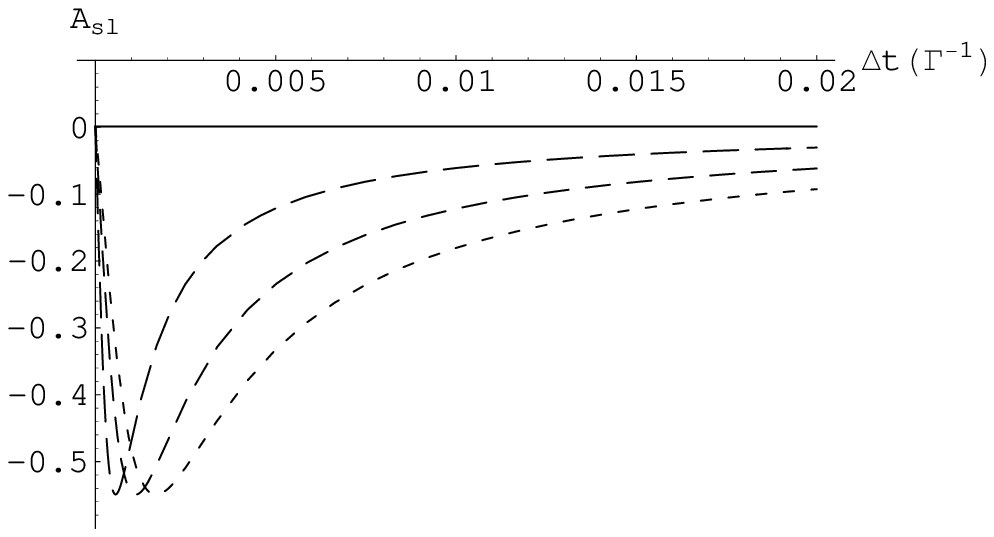}
\end{center}
\caption{\emph{Left picture}: Equal-sign dilepton charge asymmetry $A_{sl}$ for different values of $\omega = |\omega|e^{i\Omega}$, with $\Omega = 0$: $|\omega|=0$ (solid line), $|\omega|=0.0005$ (long-dashed), $|\omega|=0.001$ (medium-dashed), $|\omega|=0.0015$ (short-dashed).  When $\omega\neq0$ a peak of height $A_{sl}(peak) = 0.77 \cos(\Omega)$ appears at $\Delta t(peak)=1.12\, |\omega| \, \frac{1}{\Gamma}$,
$\Gamma = (\Gamma_1 + \Gamma_2)/2$, producing a drastic difference with the $\omega=0$ case, in particular in its time dependence.  Observe that the peak, independently of the value of $|\omega|$, can reach enhancements up to $10^3$ times the value of the asymmetry when $\omega=0$. \emph{Right picture}: as in Left picture, but for $\Omega = 3\pi/2$~\cite{bomega}.}
\label{asl}
\end{figure}
In B-factories one can look for similar $\omega$-like effects.
Although in this case, in contrast to a $\phi$-factory,  there is no particularly good channel to lead to enhancement of the sensitivity,  nevertheless one gains in statistics, and hence interesting limits may also be obtained~\cite{bomega}. The presence of a quantum-gravity induced $\omega$-effect in B systems is associated with a \emph{theoretical limitation on flavour tagging}, namely the fact that, in the absence of such effects, the knowledge that one of the two-mesons in a meson factory decays at a given time through a "flavour-specific channel'' determines unambiguously the flavour of the other meson at the same time. This is not true if intrinsic CPT Violation is present.

One of the relevant observables~\cite{bomega} is given by the CP-violating semi-leptonic decay charge asymmetry (in equal-sign dilepton channel), with the first decay $B \to X\ell^{\pm}$ being time-separated by the second decay
$B \to X'\ell^{\pm}$ by an interval $\Delta t$:
\begin{equation}
A_{sl}(\Delta t)  = \frac{I(\ell^+,\ell^+, \Delta t) - I(\ell^-,\ell^-, \Delta t)}{I(\ell^+,\ell^+, \Delta t) + I(\ell^-,\ell^-, \Delta t)} ~,
\label{asl2}
\end{equation}
where $I(\Delta t)$ denotes the relevant intensity, integrated over the time of the first decay,
$I(X \ell^\pm,X' \ell^\pm , \Delta t) = \int_0^\infty \left| \langle X \ell^\pm ,X'\ell^\pm |U(t_1) \otimes U(t_1+\Delta t) |\psi(0)\rangle \right|^2 dt_1 $, with $U(t)|B_i\rangle =e^{-i(m_i - i\Gamma_i)t}|B_i\rangle~, ~i=1,2 $ the evolution operator for mass-eigenstates states with mass $m_i$ and widths $\Gamma_i$, $i=1,2$.
One finds~\cite{bomega}:
\begin{eqnarray}
 I(X\ell^\pm, X'\ell^\pm, \Delta t) = \frac{1}{8} e^{-\Gamma \Delta t} \, |A_X|^2 |A_{X'}|^2 \, \left|\frac{(1+s_\epsilon\, \epsilon)^2-\delta^2/4}{1-\epsilon^2+\delta^2/4}\right|^2 \times \qquad ~ \ \qquad\ \quad\nonumber \\
 \qquad ~\qquad
 \begin{array}{l}
 \left\{ \Bigg[ \frac{1}{\Gamma} + a_{\omega} \frac{8\Gamma}{4\Gamma^2 + (\Delta m)^2} Re(\omega) + \frac{1}{\Gamma} |\omega|^2 \Bigg] \cosh \left(\frac{\Delta \Gamma \Delta t}{2}\right) + \right. \\
 \Bigg[ - \frac{1}{\Gamma} + b_\omega \frac{8\Gamma}{4\Gamma^2 + (\Delta m)^2}
 Re(\omega) -\frac{\Gamma}{\Gamma^2 + (\Delta m)^2} |\omega|^2 \Bigg] \cos (\Delta m \Delta t) +  \\
  \left. \Bigg[ d_\omega\frac{4 \Delta m}{4\Gamma^2 + (\Delta m)^2}  Re(\omega)  + \frac{\Delta m}{\Gamma^2 + (\Delta m)^2} |\omega|^2 \Bigg] \sin (\Delta m \Delta t) \right\},
 \end{array}
 \label{basymevol}
 \end{eqnarray}
 to leading order in small $\omega$ and conventional CPT-violating $\delta$ effects. The
 quantity $\epsilon $ denotes the CP violation parameter, while $\Gamma = \frac{\Gamma_1 + \Gamma_2}{2}$. The numerical coefficients $s_\epsilon,\ a_\omega,\ b_\omega$ and $d_\omega$ for the different decays are given in ref.~\cite{bomega} and are either $1$ or $-1$.

In the absence of $\omega$-effects, the intensity at equal decay times vanishes, $I_{\rm sl}(\ell^{\pm},\ell^{\pm},\Delta t=0) = 0$, whilst in the presence of a complex $\omega=|\omega|e^{i\Omega}$,
\begin{equation}
I_{\rm sl}(\ell^{\pm},\ell^{\pm},\Delta t=0) \sim |\omega |^2~.
\label{intbfact}
\end{equation}
In such a case, the function $A_{sl}(\Delta t)$ vs. $\Delta t$,
in the region $\frac{1}{\Gamma} \le \Delta t \le \frac{2\pi}{\Delta m}$,
exhibits a peak, whose position depends on $|\omega|$, while the shape of the curve itself depends on the phase $\Omega $ (c.f. figure~\ref{asl}). On the other hand,
the peak structure repeats itself at larger times $\Delta m \Delta t \simeq 2\pi \simeq 8.2 \Gamma^{-1} \Delta m$, due to the
periodicity of (\ref{basymevol}) in $\Delta m \Delta t$ (the reader should notice that
for small $\Delta \Gamma$, as appropriate for B-systems, the terms cosh$(\Delta \Gamma \Delta t/2)$ in (\ref{basymevol}) are approximately constant).

The analysis of Alvarez, Bernabeu and Nebot in Ref.~\cite{bomega}, using the above charge asymmetry method and comparing with currently available experimental data from B-factories (BaBar and Belle Collaborations) on $A_{sl}$ in the region $0.8\Gamma^{-1} < \Delta t < 10 \Gamma^{-1}$, i.e.:
\begin{equation}
A_{sl}^{\rm exp} = 0.0019 \pm 0.0105 ~,
\label{bresults}
\end{equation}
resulted in the following bounds for the $\omega$-effect:
\begin{equation}
 -0.0084 \le {\rm Re}(\omega) \le 0.0100 \qquad {\rm at~ 95\% ~C.L.}
 \label{bfinalresult}
 \end{equation}
It is understood that the current experimental
limits give the charge asymmetry as constant, since the relevant analysis has been done in the absence of $\omega$-effects that are responsible for the induced time dependence of this quantity. This has been properly taken into account in the relevant works of Ref.~\cite{bomega}, when placing bounds.

Unfortunately at large $\Delta t$ regions, near the points where the peak structure
 of $A_{sl}$ repeats itself, the amount of events is suppressed by a factor $e^{-8.2} \sim 10^{-4}$ and hence, currently, such measurements cannot give any useful complementary information on $\omega$-effects. At the time when these proceedings were written the author is not aware of any further developments regarding such measurements.

Before closing this subsection we would like to point out that an observation of the $\omega$-effect in both the $\phi$ and B-factories could also provide in principle an independent test of Lorentz symmetry properties of the intrinsic CPT Violation, namely whether the effect respects Lorentz symmetry.
This is simply because, although the $\Phi$ particle in neutral Kaon factories is produced at rest,
the corresponding $\Upsilon$ state in B-factories is boosted, and hence there is a frame change between the two experiments. If the quantum gravity $\omega$-effect is locally Lorentz violating, as it may happen in the models of~\cite{bms}, then a difference in value between the two
experiments should be expected, due to frame-dependence, that is dependence on the relative Lorentz factor $\gamma_L$.

\subsubsection{Disentangling the $\omega$-Effect from
the C(even)  Background and Decoherent
Evolution Effects}

When interpretating experimental results on delicate violations of CPT
symmetry, it is important
to disentangle (possible) genuine effects from those
due to ordinary physics.
Such a situation arises in connection with the $\omega$-effect,
that must be disentangled from the C(even) background
characterizing the decay products in a
$\phi$-factory~\cite{dunietz}.  The C(even) background $e^+e^- \Rightarrow 2\gamma \Rightarrow K^0
{\overline K}^0$ leads to states of the form
\begin{eqnarray*}
|b> = |K^0 {\overline K}^0 (C({\rm even}))> =
\frac{1}{\sqrt{2}}\left(K^0({\vec k}) {\overline K}^0 (-{\vec k})
+  {\overline K}^0({\vec k}) K^0 (-{\vec k}) \right)~, \nonumber
\end{eqnarray*} which at first sight mimic the $\omega$-effect, as such states would
also produce contamination by terms $K_SK_S,~K_LK_L$.

Closer inspection reveals, however,
that the two types of effects can be clearly disentangled
experimentally~\cite{bmp,poland} on two accounts:
(i) the expected magnitude of the two effects,
in view of the the above-described estimates in QG models and the unitarity bounds that suppress the `fake' (C(even)-background) effect~\cite{dafne2},
and (ii) the
different way the genuine QG-induced $\omega$-effect interferes with the
the C(odd) background~\cite{bmp}.

Finally, we remark that it is also possible to disentangle~\cite{bmp}
the $\omega$-effect from decoherent evolution
effects~\cite{bmp}, due to their different structure.
For instance, the experimental disentanglement of $\omega$ from the decoherence
parameter $\gamma$ in completely-positive QG-Lindblad models of entangled neutral mesons~\cite{ehns,lopez,benatti} is possible as a result of different symmetry properties and different structures generated by the time evolution
of the pertinent terms.

\subsubsection{Microscopic Models for the $\omega$-Effect and Order-of-Magnitude
Estimates}

For future experimental searches for the $\omega$-effect it is
important to estimate its expected order of magnitude, at least  in
some models of foam.

A specific model is that of the D-particle foam~\cite{emw,msdark,bms},
discussed already in connection with the stochastic
metric-fluctuation approach to decoherence. An important feature for
the appearance of an $\omega$-like effect is that, during each
scattering with a D-particle defect, there is (momentary) capture of
the string state (representing matter) by the defect, and a possible
\emph{change} in phase and \emph{flavour} for the particle state emerging from
such a capture (see Fig.~\ref{drecoil}).

 The induced metric distortions,
including such flavour changes for the emergent post-recoil matter
state, are:
\begin{eqnarray}
&& g^{00} =\left( -1+r_{4}\right) \mathsf{1}~, \quad
g^{01}=g^{10}=r_{0}\mathsf{1}+ r_{1}\sigma_{1}+ r_{2}\sigma_{2}
+r_{3}\sigma_{3}, \quad
 g^{11} =\left( 1+r_{5}\right) \mathsf{1} \end{eqnarray} where
the $\sigma_i$ are Pauli matrices, referring to flavour space.

The target-space metric state is the density matrix $\rho_{\mathrm{grav}}$
defined in (\ref{gravdensity})~\cite{bms}, with the same assumptions for
the parameters $r_{\mu}$ stated there.
The order of magnitude  of the metric elements $g_{0i}\simeq
\overline{v}_{i,rec} \propto g_s\frac{\Delta p_{i}}{M_{s}} $, where
$\Delta p_{i}\sim {\tilde \xi} p_i $ is the momentum transfer during
the scattering of the particle probe (kaon) with the D-particle
defect, $g_{s}<1$ is the string coupling, assumed weak, and $M_{s}$
is the string scale, which in the modern approach to string/brane
theory is not necessarily identified with the four-dimensional
Planck scale, and is left as a phenomenological parameter to be
constrained by experiment.

To estimate the order of magnitude of the $\omega$-effect we
construct the gravitationally-dressed initial entangled state using
stationary perturbation theory for degenerate states~\cite{bms}, the
degeneracy being provided by the CP-violating effects. As
Hamiltonian function we use

{\begin{eqnarray}
\widehat{H}=g^{01}\left( g^{00}\right)
^{-1}\widehat{k}-\left( g^{00}\right) ^{-1}\sqrt{\left(
g^{01}\right) ^{2}{k}^{2}-g^{00}\left( g^{11}k^{2}+m^{2}\right)  }
\end{eqnarray}}describing propagation in the above-described
stochastically-fluctuating space-time. To leading order in the
variables $r$ the interaction Hamiltonian reads:
\begin{equation}
\widehat{H_{I}} = -\left(  { r_{1} \sigma_{1} + r_{2} \sigma_{2}}
\right) \widehat{k} \nonumber
\end{equation}
with the notation {\small $\left| K_{L}\right\rangle =\left|
\uparrow\right\rangle~, \quad \left| K_{S}\right\rangle =\left|
\downarrow\right\rangle .$} The gravitationally-dressed initial states
then can be constructed using stationary perturbation theory:

{\begin{eqnarray} \left|  k^{\left(  i\right)
},\downarrow\right\rangle _{QG}^{\left( i\right)  } =  \left| k^{\left(
i\right) },\downarrow\right\rangle ^{\left( i\right)  } + \left| k^{\left(
i\right) },\uparrow\right\rangle ^{\left( i\right) } \alpha^{\left(
i\right)  }~,
\end{eqnarray}}where {\small $ \alpha^{\left(  i\right)  }= \frac{^{\left( i\right)
}\left\langle \uparrow, k^{\left( i\right) }\right| \widehat{H_{I}}\left|
k^{\left(  i\right)  }, \downarrow\right\rangle ^{\left(  i\right)
}}{E_{2} - E_{1}} $}. For $\left|  { k^{\left(  i\right) }, \uparrow}
\right\rangle ^{\left( i \right)  } $  the dressed state is obtained by
$\left| \downarrow\right\rangle \leftrightarrow\left|
\uparrow\right\rangle $ and $\alpha\to\beta$ where  {\small $
\beta^{\left( i\right) }= \frac{^{\left( i\right)  }\left\langle
\downarrow, k^{\left( i\right)  }\right| \widehat{H_{I}}\left| k^{\left(
i\right)  }, \uparrow\right\rangle ^{\left(  i\right) }}{E_{1} - E_{2}}$}.

The totally antisymmetric ``gravitationally-dressed'' state of two mesons
(kaons or B-mesons \emph{etc}.) is then~\cite{bms}:

{
\begin{eqnarray}
\begin{array}
[c]{l}%
\left|  {k, \uparrow} \right\rangle _{QG}^{\left(  1 \right)  }
\left|  { - k, \downarrow} \right\rangle _{QG}^{\left(  2 \right)  }
- \left|  {k, \downarrow} \right\rangle _{QG}^{\left(  1 \right)  }
\left|  { - k, \uparrow} \right\rangle _{QG}^{\left(  2 \right)  }
= \\
\left|  {k, \uparrow} \right\rangle ^{\left(  1 \right) } \left|
{ - k, \downarrow} \right\rangle ^{\left(  2 \right)  } - \left| {k,
\downarrow} \right\rangle ^{\left(  1 \right)  } \left| { - k,
\uparrow} \right\rangle
^{\left(  2 \right)  }\\
+  \left|  {k, \downarrow} \right\rangle ^{\left(  1 \right) }
\left|  { - k, \downarrow} \right\rangle ^{\left(  2 \right)  }
\left(  {\beta^{\left(  1 \right)  } - \beta^{\left(  2 \right)  } }
\right) + \\
\left|  {k, \uparrow} \right\rangle ^{\left( 1 \right)
} \left|  { - k, \uparrow} \right\rangle ^{\left(  2 \right)  }
\left( {\alpha^{\left(  2 \right)  } - \alpha^{\left(
1 \right)  } } \right) \\
 + \beta^{\left(  1 \right)  } \alpha^{\left(  2 \right) }
\left|  {k, \downarrow} \right\rangle ^{\left(  1 \right) } \left| {
- k, \uparrow} \right\rangle ^{\left(  2 \right)  } - \alpha^{\left(
1 \right)  } \beta^{\left( 2 \right)  } \left|  {k, \uparrow}
\right\rangle ^{\left(  1
\right)  } \left|  { - k, \downarrow} \right\rangle ^{\left(  2 \right)  }~.\\
\label{entangl}%
\end{array}\end{eqnarray}}Notice here
that, for our order-of-magnitude estimates, it suffices to assume that the
initial entangled state of kaons is a pure state. In practice, due to the
omnipresence of foam, this may not be entirely true, but this should not
affect our order-of-magnitude estimates based on such an assumption.

With these remarks in mind we then write for the initial state of
two mesons, say, for concreteness, the two kaons after the $\phi$ decay:
{\small \begin{eqnarray}\label{entkaon}
&&\left| \psi\right\rangle = \left|  k,\uparrow\right\rangle
^{\left( 1\right) }\left| -k,\downarrow\right\rangle ^{\left(
2\right) }-\left| k,\downarrow\right\rangle ^{\left(  1\right)
}\left|
-k,\uparrow \right\rangle ^{\left(  2\right)  }+
 \xi \left| k,\uparrow\right\rangle ^{\left( 1\right) }\left|
-k,\uparrow\right\rangle ^{\left(  2\right) }+  \xi^{\prime} \left|
k,\downarrow\right\rangle ^{\left( 1\right) }\left|
-k,\downarrow\right\rangle ^{\left( 2\right)  }~,
\end{eqnarray}}where for $r_{i} \propto\delta_{i1} $ we have
$\xi= \xi^{\prime}$, that is strangeness violation, whilst for
$r_{i} \propto\delta_{i2}$ $\longrightarrow $ $\xi= -\xi^{\prime}$)
(since $\alpha^{\left( i \right) } = \beta^{\left( i \right) } )$ we
obtain a strangeness conserving $\omega$-effect.

Upon averaging the density matrix over $r_{i}$, only the
$|\omega|^{2}$ terms survive: {\small \begin{eqnarray}
|\omega|^{2} = \mathcal{O}\left(  \frac{1}{(E_{1} - E_{2})^2}
(\langle \downarrow, k |H_{I} |k, \uparrow\rangle)^{2} \right)  \sim
\frac{\Delta_{2} k^{2}}{(m_{1} - m_{2})^{2}} \label{omegaestim}
\end{eqnarray}}for momenta of order of the rest energies, as is the case of a
$\phi$ factory. We arrived at this result by assuming independent variances between the particle species ``1'' and ``2'', representing the $K_L$ and $K_S$ states respectively.
In this sense, the variance $\Delta_2$ represents an appropriate difference of variances
for the left and the right particles after the initial $\phi$-meson decay.

Recalling that in the recoil D-particle model under consideration we
have~\cite{emn,bms} $\Delta_{2} = {\tilde \xi}^{2} k^{2}/M_{P}^{2}$,
we obtain the following order of magnitude estimate of the $\omega$
effect: {\small \begin{eqnarray} |\omega|^{2} \sim\frac{{\tilde
\xi}^{2} k^{4}}{M_{P}^{2} (m_{1} - m_{2})^{2}}~. \label{orderomega}
\end{eqnarray}}For neutral kaons with momenta of the order of the rest energies
$|\omega| \sim10^{-4} |{\tilde \xi}|$.
The parameter $\tilde{\xi}$ is purely phenomenological at this stage, and its magnitude depends on the details of the microscopic model of foam.
For $1 >
{\tilde\xi}\ge 10^{-2}$ this is not far below the sensitivity of current
facilities, such as KLOE at DA$\Phi$NE~\cite{dafne2}. In fact, the KLOE experiment
has just released the latest measurement of the $\omega$
parameter, as reported by A. DiDomenico in this conference~\cite{adidomenico}: \begin{eqnarray}\label{kloe} &&
{\rm Re}(\omega) = \left(
-1.6^{+3.0}_{-2.1} \pm 0.4\right)\times 10^{-4}~, \quad
{\rm Im}(\omega) = \left( -1.7^{+3.3}_{-3.0} \pm 1.2\right)\times
10^{-4}~, \nonumber \\ && |\omega | < ~1.0 \times 10^{-3}~~{\rm at~95~\%~C.L.}\end{eqnarray}
One can constrain the $\omega$
parameter (or, in the context of the above specific model, the
momentum-transfer parameter ${\tilde \xi}$) significantly in
upgraded facilities. For instance, there are the following
perspectives for KLOE-2 at (the upgraded)
DA$\Phi$NE-2~\cite{adidomenico}:
\begin{equation}\label{prospects}
{\rm Re}(\omega),~ {\rm Im}(\omega) \longrightarrow 2 \times 10^{-5}~.
\end{equation}

In some detailed string/brane models of quantum D-particle foam,
such as the ones in ref.~\cite{emw,emnnewuncert},
the phenomenological parameter ${\tilde \xi}$ can be computed explicitly, and in fact it is found to depend on the details of the model, such as masses and momenta.
Indeed, in such models excitations of the Standard model are represented as
open strings with their ends attached on the brane worlds, while
the interactions with the foam involve capture of the open string by the D-particle
(c.f. fig.~\ref{drecoil}). During the capture process, the open string excitations
are represented by stretched strings between the D-particle and the brane world.
Only electrically neutral excitations (or those in the Cartan subalgebra of the Standard Model group) are interacting non-trivially with the D-particle foam~\cite{emnnewuncert}.
In the case of neutral mesons, considered above, it is essentially some of the gluons (in the Cartan subalgebra of the SU(3) x SU(2) xU(1) group) that can exhibit such interactions. The presence of strong interactions inside the meson ``bag'' complicates the estimates, as mentioned above, but one may make the simplifying assumption that
the entire meson can be viewed as an elementary electrically neutral entity when interacting with the foam.

Some remarks are due at this point. During that process ``flavour changes'' also occur, in the sense of oscillations
between $K_L$,$K_S$ states, say, in the concrete example of neutral Kaons.
If such flavour oscillations were due to the capture  of  matter states by the D-particle defect, during the initial decay of the $\phi$ meson, then
microscopically these would be due  to the transmutation of the quark constituents of $K^0$ and ${\overline K}^0$ into the respective antiquarks.
However, since in our model of foam, the electrically charged quark constituents of Kaons are not captured by the D-particle defect~\cite{emnnewuncert}, due to electric charge conservation, the D-particle-induced flavour oscillations are absent~\footnote{This is not the case in neutrinos, where such
D-foam induced flavour oscillations may exist. However, whether entangled states of neutrinos do exist in situations of practical usefulness, that can be used in a similar fashion to constrain the $\omega$-effect, is unclear to us at present.}.

Let $p^0$ be the energy of the incident open string(s) representing collectively the Kaon
as it hits the D-particle and splits into two open strings stretched between the latter and the brane world (c.f. \ref{drecoil}).
Then, the total energy available will be given by
$p^0 $. This energy will lead to the formation of an intermediate string state that oscillates $N$ times, growing in size from zero length till a maximum length $L$, and back to zero size, with the free end of the string on the brane world moving with the speed of light. This leads to~\cite{emnnewuncert,toumbas}:
\begin{equation}
 p^0 = \frac{L}{\alpha '} + \frac{N}{L^2}
\label{totener}
\end{equation}
Minimising the right-hand-side with respect to $L$ we obtain
the maximum $L$, as $L_{\rm max} \sim \alpha ' p^0 $ and from this the time uncertainty, which expresses the duration $t_c$ of the capture process:
\begin{equation}\label{capturetime}
   t_c \sim \alpha ' p^0
\end{equation}
As discussed in some detail in \cite{tsallis}, during the capture process, the quantum fluctuations of the recoil velocity of the D-particle can be estimated by summing up world-sheet higher-genus surfaces. Such a resummation cannot be performed in a closed form in a supersymmetric world sheet theory, unfortunately. However, one can get an estimate
of the quantum fluctuations of the recoil velocity, by performing such a resummation to leading order in the so-called modular infinities, in a bosonic $\sigma$-model in the background of a recoiling D-particle~\cite{tsallis}. In doing so, one obtains that
the quantum recoil velocity of the D-particle, fluctuates about a zero average, with a Gaussian distribution with variance over the duration of the capture~\cite{tsallis}:
\begin{equation}\label{variancerec}
  \Delta_2^{(i)} \sim g_s^2 (t_c^{(i)})^2~,
\end{equation}
where the superscript $(i)$ denotes particle species, $g_s$ is the (weak, $g_s < 1$) string coupling and the capture time $t_c$ is given by (\ref{capturetime}).

The analysis of \cite{bms,tsallis}, then, described briefly above, gives the $\omega$-like terms in the entangled state of the two Kaons (\ref{entkaon}) as the sum
of the appropriate coefficients over particle species ``1'' and   ``2'' (representing the $K_L$ and $K_S$ states respectively). In such a case, then,
assuming the presence of a D-particle defect during the decay of the initial
$\phi$-meson, which captures the relevant string excitations representing
the gluon constituents of the Kaons (since the electrically charged quarks are not supposed to interact individually with the defect), and
taking into account (\ref{capturetime}) and (\ref{variancerec}), we can estimate~\cite{tsallis}:
\begin{equation}\label{finalomega}
 |\omega|^2 \sim \left(\Delta_2^{(1)} + \Delta_2^{(2)}\right)\frac{k^2}{(E_1 - E_2)^2} \sim ~\frac{g_s^2}{M_s^2}\frac{\left(m_1^2 +  m_2^2\right)}{|m_1 - m_2|^2}k^2~,
  \end{equation}
where $M_s=1/\ell_s$ is the string mass scale, and, as appropriate for Da$\Phi$NE $\phi$-factory~\cite{dafne2}, we assumed non relativistic dispersion relations for kaons, with $E_{(i)} \sim m^{(i)} + {\cal O}(\frac{k^2}{{m^{(i)}}^2})$, for each particle species (on each side of the detector)~\footnote{Above, we have used the fact that
the variances $\Delta_2^{(i)}$ for each Kaon constituent state $i=1,2$ in the initially entangled state are proportional to the individual Kaon energies, (\ref{variancerec}),
(\ref{capturetime}). This stems from the plausible assumption, that, in the initial state after the $\phi$-meson decay, the
appropriate open string excitations in each of the two neutral kaons in our D-brane model (corresponding to electrically neutral components of the Kaon ``bag'', \emph{i.e.} gluons~\cite{emnnewuncert}) interact separately with the D-particles. On the other hand, it could be that the D-particle interacts with the composite initial entangled Kaon state as a whole. Then, the capture time would be proportional to the total energy of both kaons, and hence the variances would assume a common value for both particles, \emph{i.e}. $\Delta_2^{(i)} \sim \frac{g_s^2}{M_s^2}\left( m_1^2 + m_2^2 + {\rm momentum~dependent~parts} \right)$, $i=1,2$. In such a case, the overall estimate for the $\omega$ amplitude squared will be given by the right-hand-side of (\ref{finalomega}) multiplied by a factor of 2. This will not affect our order of magnitude considerations here. Such issues can only be resolved when a fully microscopic understanding of the interactions of the Kaon constituents with the D-particle foam, taking proper account of the strong-interaction effects, becomes available.}.

Compared to the generic relation (\ref{orderomega}), we then observe that the r\^ole of the
Planck Mass is played here by the D-particle mass~\cite{kmw}, $M_s/g_s$,
while the parameter ${\tilde \xi}$ is of order:
\begin{equation}\label{xiorder}
|{\tilde \xi} |^2 \sim \frac{m_1^2 + m_2^2}{k^2}~,
\end{equation}
and thus it depends on the details of the system, such as masses and momenta.

For neutral Kaons in a $\phi$ factory, we have
\begin{equation}
\Delta m = m_{L} - m_{S} \sim3.48 \times10^{-15}~\mathrm{GeV}~. \label{deltam}%
\end{equation}
and the momenta are of order $1$ GeV.
The result (\ref{finalomega}), then implies the estimate
\begin{equation}\label{dfoamomega}
|\omega|_{\rm D-foam} = {\cal O}\left( 10^{-5}\right)~.
\end{equation}
In the sensitive $\eta^{+-}$ bi-pion decay channel, this leads to effects enhanced by three
orders of magnitude, as a result of the fact that terms containing $|\omega|$
always appear in the corresponding observables~\cite{bmp} (c.f. (\ref{omegevoldet}), (\ref{omegevoldet2}), (\ref{omegevoldet3})) in the
form $|\omega|/|\eta^{+-}|$, and the CP-violating parameter
$|\eta^{+-}|\sim 10^{-3}$. At present, the estimate (\ref{dfoamomega}) is still some two
orders of magnitude away from the current bounds (\ref{kloe}) of the $\omega$-effect
by the KLOE collaboration at DA$\Phi$NE~\cite{adidomenico}, giving $|\omega| < 10^{-3}$,
but it is within the projected sensitivity of the proposed upgrades (c.f. (\ref{prospects})).

The above estimate is valid for non-relativistic meson states, where each meson has been treated as a structureless entity when considering its interaction with the D-foam.
As already mentioned, the inclusion of details of the Kaons strongly-interacting substructure may affect the results.

Let us now mention that $\omega$-like effects can also be generated by
the Hamiltonian evolution of the system as a result of gravitational
medium interactions. To this end, let us consider the Hamiltonian
evolution in our stochastically-fluctuating D-particle-recoil
distorted space-times, {\scriptsize  $ \left| \psi\left(  t\right)
\right\rangle =\exp\left[ -i\left( \widehat
{H}^{(1)}+\widehat{H}^{\left( 2\right) }\right)
\frac{t}{\hbar}\right] \left| \psi\right\rangle$}.

Assuming for simplicity $\xi = \xi^\prime = 0$, it is easy to
see~\cite{bms} that the time-evolved state of two kaons contains
strangeness-conserving $\omega$-terms:
\begin{eqnarray*}  \left|  \psi\left(  t\right)  \right\rangle \sim
e^{-i\left(  \lambda_{0}^{\left(  1\right)  }+
\lambda_{0}^{\left(  2\right)  }\right)  t}%
 \varpi\left(  t\right) \times
\left\{  \left| k,\uparrow\right\rangle ^{\left( 1\right) }\left|
-k,\uparrow\right\rangle ^{\left( 2\right)  }-\left|
k,\downarrow\right\rangle ^{\left(  1\right) }\left|  -k,\downarrow
\right\rangle ^{\left(  2\right)  }\right\}~. \nonumber
\end{eqnarray*}
The quantity $\varpi (t)$ obtained within this specific model is purely imaginary,
\begin{eqnarray*} {\cal O}\left(\varpi\right)  \simeq
i\frac{2\Delta_{1}^{\frac{1}{2}}k}{\left(
k^{2}+m_{1}^{2}\right)^{\frac{1}{2}}-\left(
k^{2}+m_{2}^{2}\right)^{\frac{1}{2}}}\times \cos\left( \left| \lambda^{\left(  1\right) }\right| t\right)
\sin\left( \left| \lambda^{\left(  1\right) }\right| t\right) =
\varpi_{0}\sin\left( 2\left|  \lambda^{\left( 1\right) }\right|
t\right), \nonumber
\end{eqnarray*} with $\Delta_{1}^{1/2}\sim\left| {\tilde \xi}\right| \frac{\left| k\right|
}{M_{P}}$,
$\varpi_{0}\equiv\frac{\Delta_{1}^{\frac{1}{2}}%
k}{\left(  k^{2}+m_{1}^{2}\right)  ^{\frac{1}{2}}-\left(  k^{2}+m_{2}%
^{2}\right)  ^{\frac{1}{2}}}$, $\left|  \lambda^{\left(  1\right)
}\right|  \sim\left( 1+\Delta_{4}^{\frac{1}{2}}\right)
\sqrt{\chi_{1}^{2}+\chi_{3}^{2}}$,~ $\chi_{3}\sim\left(
k^{2}+m_{1}^{2}\right)  ^{\frac{1}{2}}-\left(
k^{2}+m_{2}^{2}\right) ^{\frac{1}{2}}$.

It is important to notice the time dependence of the medium-generated
effect. It is also interesting to observe that, if in the initial state
we have  a strangeness-conserving (-violating) combination,
$\xi = -\xi^\prime $  ($\xi = \xi^\prime $),
then the time
evolution generates time-dependent strangeness-violating
(-conserving $\omega$-) imaginary effects.

The above description of medium effects using Hamiltonian evolution is
approximate, but suffices for the purposes of obtaining order-of-magnitude
estimates for the relevant parameters. In the complete description of the
above model  there is of course decoherence~\cite{bms,emn}, which affects
the evolution and induces mixed states for kaons. A complete analysis of
both effects, $\omega$-like and decoherence in entangled neutral kaons of
a $\phi$-factory, has already been carried out~\cite{bmp}, with the upshot
that the various effects can be disentangled experimentally, at least in
principle.

Finally, as the analysis of
\cite{bms,disfoam} demonstrates, no $\omega$-like effects are generated by
thermal bath-like (rotationally-invariant, isotropic) space-time
foam situations, argued to simulate the QG
environment in some models \cite{garay}. In this
way, the potential observation of an $\omega$-like effect in
EPR-correlated meson states would in principle distinguish various
types of space-time foam.

\section{Conclusions and Outlook \label{sec:5}}

In this review we have outlined several aspects of decoherence-induced CPTV and
the corresponding experiments.
We described the interesting and challenging precision tests
that can be performed using kaon systems, especially $\phi$-meson factories,
where some unique ($\omega$) effects on EPR-correlation modifications, associated with the ill-defined nature of CPT
operator in decoherent QG, may be in place.

We have presented sufficient theoretical motivation and
estimates of the associated effects to support the case that
testing QG experimentally at present
facilities may turn out to be a worthwhile endeavour.
In fact, as we have argued, CPTV may be a real feature of QG,
that can be tested and observed, if true, in the foreseeable future.
Indeed, as we have seen, some theoretical (string-inspired) models of
space-time foam predict $\omega$-like effects of an order of
magnitude that is already well within the reach of the next upgrade of
$\phi$-factories, such as DA$\Phi$NE-2.
Neutrino systems is another extremely sensitive probe of particle physics
models which can already falsify several models or place stringent
bounds in others.

We would also like to mention at this stage that QG-decoherence effects in Cosmology may also play a r\^ole in our understanding of the Universe evolution, or even modify the current astrophysical constraints on models of particle physics, where such possible decoherent effects are ignored.
The presence of space-time foam effects could be much stronger in the Early Universe, and such strong presence could affect significantly the relevant evolution equations (Boltzmann type) for the density of species and hence the various cosmic relic abundances and the contributions to the dark matter sector of our Universe.

Therefore, the current experimental situation for QG
signals, including those for CPT Violation,  appears exciting,
and several experiments are reaching interesting regimes, where many
theoretical models can be falsified. More precision experiments are
becoming available, and many others are being designed for the
immediate future. Searching for tiny effects of this elusive theory
may at the end be very rewarding. Surprises may be around the
corner, so it is worth investing time and effort.

In particular, from the modern point of view of quantum gravity, inspired by string theory, in which large extra dimensions may exist~\cite{add,rs}, and the bulk-space string scale $M_s$ is essentially a free parameter, different from the Planck scale, and possibly as low as a few TeV, issues on quantum-gravity decoherence
acquire a new perspective. Indeed, the possibility of the formation of mini- (string-mass scale) black holes at LHC, points towards the possibility of exposing ourselves to the regime of quantum space-time foam directly. Although many of the decoherence models we have discussed here would have effects that could still be compatible with TeV-size black holes, some are not.
For instance, the $\omega$-effect theoretical estimate (\ref{orderomega}), which is based on such a string-inspired theory, would increase
drastically, in case the quantum gravity scale $M_P$ was as low as a few TeV. Reversing the logic, if effects like the $\omega$-effect are seen in the foreseeable future, then this would give information on the
size of the characteristic scale for which quantum gravity effects set in in these extra dimensional models of string theory.
It is interesting therefore to
perform detailed studies of quantum decoherence in such higher-dimensional models as well.
First steps in this direction have already been taken in \cite{anchor}, and one might expect that the launch of LHC will spark an increased research activity along this direction.

 I conclude the talk by mentioning that there are many other possible properties, apart from decoherence, that could characterise a quantum gravity model, which we did not discuss here. An important such possibility is that of Lorentz Symmetry Violation or modification by quantum gravity.
Others speakers in this conference have touched upon such topics, either from the point of view of terrestrial precision tests of Lorentz symmetry in, say, atomic and nuclear physics~\cite{sme} or by means of astrophysical tests of the constancy of the speed of light~\cite{manel}. I refer the reader to their contributions in this volume for details.

\section*{Acknowledgements}

This article is based on a plenary talk given by the author at the DISCRETE '08 \emph{Symposium
on Prospects in the Physics of Discrete Symmetries}, IFIC,
Valencia (Spain), December 11-16 2008. The work is partially supported by the European Union through the Marie Curie Research and Training Network \emph{UniverseNet}
(MRTN-2006-035863).

\end{document}